\documentclass[11pt]{article}

\usepackage{amsmath}
\usepackage{amssymb}
\usepackage{amsthm}
\usepackage{amsfonts}
\usepackage{amstext}
\usepackage{amsopn}
\usepackage{amsxtra}
\usepackage{graphicx}
\usepackage[latin1]{inputenc}
\usepackage{mathrsfs}
\usepackage{dsfont}
\usepackage{xr}
\newcommand{\refII}[1]{{\rm II.}\ref{#1}}
\newcommand\R{{\ensuremath {\mathbb R} }}
\newcommand\C{{\ensuremath {\mathbb C} }}

\newcommand\Z{{\ensuremath {\mathbb Z} }}

\newcommand\1{{\ensuremath {\mathds 1} }}

\renewcommand\phi{\varphi}

\newcommand{\cP}{\mathcal{P}}

\newcommand{\cB}{\mathcal{B}}

\newcommand{\cM}{\mathcal M}

\newcommand{\cR}{\mathcal R}

\newcommand{\cE}{\mathcal{E}}

\newcommand\ii{{\ensuremath {\infty}}}

\newcommand{\norm}[1]{ \left| \! \left| #1 \right| \! \right| }

\newtheorem{thm}{Theorem}
\newtheorem{lemma}{Lemma}

\newtheorem{prop}[lemma]{Proposition}
\newtheorem{definition}{Definition}
\newtheorem{remark}{Remark}

\date{December 18, 2008}
\begin{document}

\title{The Thermodynamic Limit of Quantum Coulomb Systems\\ Part I. General Theory}
\author{C. HAINZL, M. LEWIN and J. P. SOLOVEJ} 

\begin{center}
  \bf \Large The Thermodynamic Limit of Quantum\\ Coulomb Systems

\medskip

Part I. General Theory
\end{center}

\medskip

\begin{center}
 \large Christian HAINZL$^a$, Mathieu LEWIN$^b$ \& Jan Philip SOLOVEJ$^c$
\end{center}

\medskip

\begin{center}
\small

 $^a$Department of Mathematics, UAB, Birmingham, AL 35294-1170, USA. 

\texttt{hainzl@math.uab.edu}

\medskip

$^b$CNRS \& Department of Mathematics UMR8088, University of Cergy-Pontoise, 2 avenue Adolphe Chauvin, 95302 Cergy-Pontoise Cedex, FRANCE. 

\texttt{Mathieu.Lewin@math.cnrs.fr}

\medskip

$^c$University of Copenhagen, Department of Mathematics, Universitetsparken 5, 2100 Copenhagen, DENMARK. 

\texttt{solovej@math.ku.dk}
\end{center}

\medskip

\begin{center}
 \it December 18, 2008
\end{center}

\medskip

\begin{abstract}
This article is the first in a series dealing with the thermodynamic properties of quantum Coulomb systems. 

In this first part, we consider a general real-valued function $E$ defined on all bounded open sets of $\R^3$. Our aim is to give sufficient conditions such that $E$ has a thermodynamic limit. This means that the limit $E(\Omega_n)|\Omega_n|^{-1}$ exists for all `regular enough' sequence $\Omega_n$ with growing volume, $|\Omega_n|\to\ii$, and is independent of the considered sequence.

The sufficient conditions presented in our work all have a clear physical interpretation. In the next paper, we show that the free energies of many different quantum Coulomb systems satisfy these assumptions, hence have a thermodynamic limit.
\end{abstract}

\bigskip

\newpage
\tableofcontents

\section*{Introduction}\addcontentsline{toc}{section}{Introduction}
Ordinary matter is composed of electrons (negatively charged) and nuclei (positively charged) interacting via Coulomb forces. The potential between two such particles of charges $z$ and $z'$ located at $x$ and $x'$ in $\R^3$ is
$$\frac{zz'}{|x-x'|}.$$
There are two difficulties which occur when trying to describe systems composed of electrons and nuclei. Both have to do with the physical problem of \emph{stability} of quantum systems.

\medskip

The first is due to the singularity of $1/|x|$ at $0$: it is necessary to explain why a particle will not rush to a particle of the opposite charge. One of the first major triumphs of the theory of quantum mechanics is the explanation it gives of the stability of the hydrogen atom (and the complete description of its spectrum) and of other microscopic quantum Coulomb systems. 

Quantum Mechanics or more precisely the uncertainty principle explains not only the stability of tiny microscopic objects, but also the stability of gigantic stellar objects such as white dwarfs. Chandrasekhar's famous theory on the stability of white dwarfs required, however, not only the
usual uncertainty principle, but also the Pauli exclusion principle for the fermionic electrons.

Both the stability of hydrogen and the stability of white dwarfs simply mean that the total energy of the system cannot be arbitrarily
negative.  If there was no such lower bound to the energy one would have a system from which it would be possible in principle to extract
an infinite amount of energy. One often refers to this kind of stability as {\it stability of the first kind} \cite{Lieb1,Lieb2}.
If we denote by $E(N)$ the ground state energy of the system under consideration, for $N$ particles, stability of the first kind can be written
\begin{equation}
 E(N)>-\ii.
\label{stability_first_kind}
\end{equation}

In proving \eqref{stability_first_kind} for Coulomb systems, a major role is played by the uncertainty principle which is mathematically expressed by  the critical Sobolev embedding $H^1(\mathbb{R}^3)\hookrightarrow L^6(\mathbb{R}^3)$. The latter allows to prove Kato's inequality 
$$\frac1{|x|}\leq \epsilon(-\Delta)+\frac{1}\epsilon,$$
for all $\epsilon>0$, which means that the Coulomb potential is controlled by the kinetic energy.

\medskip

The second issue concerns the slow decay of $1/|x|$ at infinity and this has to do with the macroscopic behavior of quantum Coulomb systems. 
It is indeed necessary to explain how a very large number of electrons and nuclei can stay bounded together to form macroscopic systems, although each particle interacts with a lot of other charged particles due to the long tail of the Coulomb interaction potential.
Whereas both the stability of atoms and the stability of white dwarfs were early triumphs of quantum mechanics it, surprisingly, took nearly forty years before the question of stability of everyday macroscopic objects was even raised (see Fisher and Ruelle \cite{fr}). The rigorous answer to the question came shortly thereafter in what came to be known as the Theorem on Stability of Matter proved first by Dyson and Lenard \cite{DL1,DL2}. 

The main question is how the lowest possible energy $E(N)$ appearing  in \eqref{stability_first_kind} depends on the (macroscopic) number $N$ of particles in the object.
More precisely, one is interested in proving a behavior of the form
\begin{equation}
 E(N)\sim_{N\to\ii}\bar{e}N.
\label{thermo_lim_N_intro}
\end{equation}
This behavior as the number of particles grows is mandatory to explain why matter does not collapse or explode in the thermodynamic limit.
Assume that \eqref{thermo_lim_N_intro} does not hold and that for instance $E(N)\sim_{N\to\ii}cN^p$ with $p\neq1$. Then $|E(2N)-2E(N)|$ becomes very large as $N\gg1$. Depending on $p$ and the sign of the constant $c$, a very large amount of energy will be either released when two identical systems are put together, or necessary to assemble them.
The constant $\bar{e}$ in \eqref{thermo_lim_N_intro} is interpreted as the energy per particle.

Stability of Matter is itself a necessary first step towards a proof of \eqref{thermo_lim_N_intro} as it can be expressed by the lower bound:
\begin{equation}
 E(N)\geq -\kappa N.
\label{stability_of_matter_N_intro}
\end{equation}
Put differently, the lowest possible energy calculated per particle
cannot be arbitrarily negative as the number of particles increases.
This is also often referred to as {\it stability of the second kind} \cite{Lieb1,Lieb2}.

A maybe more intuitive notion of stability would be to ask for the
volume occupied by a macroscopic object. Or more precisely, what is
the volume of the object when its total energy is close to the lowest
possible. It can be shown in general that this properly defined volume is proportional to the number of particles $N$. 
Denoting by $\Omega$ a domain in $\R^3$ which is occupied by the system under consideration and by $E(\Omega)$ its (lowest possible) energy, \eqref{thermo_lim_N_intro} then reads
\begin{equation}
 E(\Omega)\sim_{|\Omega|\to\ii}\bar{e}|\Omega|
\label{thermo_lim_V_intro}
\end{equation}
where $|\Omega|$ is the volume of $\Omega$. Similarly, stability of the second kind is then expressed as
\begin{equation}
 E(\Omega)\geq -\kappa|\Omega|.
\label{stability_of_matter_V_intro}
\end{equation}

Large quantum Coulomb systems have been the object of an important investigation in the last decades and many techniques have been developed.  A result like \eqref{stability_of_matter_N_intro} (or equivalently \eqref{stability_of_matter_V_intro}) was first proved for quantum electrons and nuclei by Dyson and Lenard \cite{DL1,DL2}. 
After the original proof by Dyson and Lenard several other proofs were given.  Lieb and Thirring~\cite{LT} in particular presented an elegant and simple proof relying on an {uncertainty principle for fermions}. The different techniques and results concerning stability of matter were reviewed for instance in \cite{Lieb1,Lieb2,l-heisen,Loss,Solovej_rev}.

It is very important that the negatively charged particles (the electrons) are fermions. It was discovered by Dyson in \cite{dyson1} that the Pauli exclusion principle is essential for stability of Coulomb systems: charged bosons are alone not stable because their ground state energy satisfies $E(N)\sim -CN^{7/5}$, see \cite{dyson1,CLY1,LiSo,Solovej3}.

A result like \eqref{thermo_lim_N_intro} (or equivalently \eqref{thermo_lim_V_intro}) was first proved for quantum Coulomb systems by Lieb and Lebowitz in \cite{LL} for a system containing electrons and nuclei both considered as dynamic quantum particles, hence in particular invariant under rotations. Later Fefferman gave a different proof \cite{F} for the case where the nuclei are fixed particles placed on a lattice, a system which is not invariant under rotations.

Instead of the ground state energy, one can similarly consider the free energy $F(\Omega,\beta,\mu)$ at temperature $T=1/\beta$ and chemical potential $\mu$. A precise definition of this quantity will be provided for the three examples that we shall consider in \cite{2}. One is interested in proving the equivalent of \eqref{thermo_lim_V_intro}
\begin{equation}
 F(\Omega,\beta,\mu)\sim_{|\Omega|\to\ii}\bar{f}(\beta,\mu)|\Omega|
\label{thermo_lim_V_intro2}
\end{equation}
where $\bar{f}(\beta,\mu)$ is the free energy per unit volume. The pressure is then given by $p(\beta,\mu)=-\beta \bar{f}(\beta,\mu)$.

\bigskip

In this work, we provide a new insight in the study of the thermodynamic limit of quantum systems, by giving a general proof of \eqref{thermo_lim_V_intro} or \eqref{thermo_lim_V_intro2} which can be applied to many different quantum systems including those studied by Lieb and Lebowitz in \cite{LL} or Fefferman in \cite{F}. Our goal was to identify the main physical properties of the free energy which are sufficient to prove the existence of the thermodynamic limit. For this reason, this first paper will be dedicated to the study of an abstract energy
$$\Omega\mapsto E(\Omega)$$
defined on all open bounded subsets of $\R^3$ with values in $\R$, for which the limit
\begin{equation}
 \lim_{n\to\ii}\frac{E(\Omega_n)}{|\Omega_n|}=\bar e
\label{thermo_limit_intro_general}
\end{equation}
holds as $|\Omega_n|\to\ii$. In principle, $E$ can represent the ground state energy or the free energy at temperature $T>0$ of any quantum (or even classical) system. 

We shall apply our general framework to several quantum Coulomb systems in another paper\footnote{Equations or results with reference $n$ in the second paper \cite{2} will be denoted as ${\rm II}.n$.} \cite{2}. More precisely, we shall treat in \cite{2} three different systems:
\begin{itemize}
\item the quantum crystal for which the nuclei are fixed particles on a lattice, with a possible local rearrangement and local defects;
\item dynamic quantum nuclei and electrons in a periodic magnetic field;
\item optimized classical nuclei with quantum electrons.
\end{itemize}

Like in previous works, our method consists in first showing the existence of the limit \eqref{thermo_limit_intro_general} for a specific domain $\triangle$ which is dilated (and possibly rotated and translated). Usually $\triangle$ is chosen to be a ball, a cube or a tetrahedron. In the applications \cite{2} we always use a tetrahedron as we shall heavily rely on an inequality proved by Graf and Schenker \cite{GS} for this type of domains, see Proposition \refII{Prop_Graf_Schenker}.

The second step consists in showing the existence of the limit \eqref{thermo_limit_intro_general} for any (reasonable) sequence of domains $\{\Omega_n\}$ such that $|\Omega_n|\to\ii$. This is important as in principle the limit could depend on the chosen sequence, a fact that we want to exclude for our systems. Here we do not specify what a ``reasonable'' sequence is but some properties will need to be assumed to ensure that boundary effects always stay negligible. It is not very surprising that proving the existence of the limit for any sequence $\{\Omega_n\}$ is much more complicated (hence requires more assumptions on the energy $E$) than for a simple sequence made of the reference set $\triangle$. For this reason, we have divided our results in two parts and first treat the simpler case of $\triangle$.

It is to be noticed that our method (relying on the Graf-Schenker inequality) is primarily devoted to the study of quantum systems interacting through Coulomb forces. It might be applicable to other interactions but we shall not address this question here.

\medskip

We now describe vaguely the assumptions which we found to be sufficient on the function $E$ to prove the existence of the limit \eqref{thermo_limit_intro_general}. Of course we shall give more details later on. 

\begin{figure}[h]
\centering
\includegraphics[width=6cm]{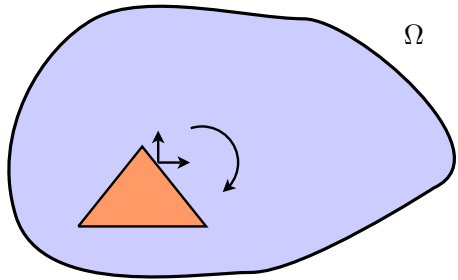}
\caption{Assumption \textbf{(A5)}: the energy of $\Omega$ is greater than the energy per unit volume of the reference set $\triangle$, averaged over all possible translations and rotations inside $\Omega$.}
\label{fig:A5}
\end{figure}

To prove the existence of the limit for sequences made of the reference set $\triangle$, we need five main assumptions. In principle, $\triangle$ can be any convex bounded open set. The first is the normalization condition:
\begin{description}
\item[(A1)]  $E(\emptyset)=0$.
\end{description}
The second important assumption on $E$ is the \emph{stability of matter}:
\begin{description}
\item[(A2)] $\forall\Omega\in\cM,\quad E(\Omega)\geq -\kappa|\Omega|$.
\end{description}
as we explained before. In the following $\cM$ is by definition the set of all bounded open subsets of $\R^3$, whereas $\cR$ will be a subclass of `regular' sets in a sense which will be made precise later. Determining the correct regularity assumption is indeed a subtle task which we do not explain more in this introduction. We also need some \emph{translation invariance property}: we assume that the following limit exists for all $\Omega\in\cR$:
\begin{description}
\item[(A3)]  $\displaystyle \lim_{L\to\ii}L^{-3}\int_{|u|\leq L}\frac{E(\Omega+u)}{|\Omega|}\,du.$
\end{description}
Examples are given by fully-translation invariant systems, or periodic systems. But we can also treat perturbations of those.
The following \emph{continuity} property will play an important role: 
\begin{description}
\item[(A4)] $\forall \Omega',\Omega\in\cR\text{ with } \Omega'\subset\Omega,\quad E(\Omega)\leq E(\Omega')+\kappa|\Omega\setminus\Omega'|+\text{error}.$
\end{description}
It essentially says that a small decrease of $\Omega$ will not decrease too much the energy. A similar property was used and proved in the crystal case by Fefferman \cite[Lemma 2]{F}. In the above formula (and below), `error' describes an error term which will be precised later. Indeed, as we will see the specific form of the error terms will play an important role in the proof of the existence of the thermodynamic limit.
Finally, our method is mainly based on the following inequality
\begin{description}
\item[(A5)] $\displaystyle\forall\Omega\in\cM,\quad E(\Omega)\geq \frac{1}{|\triangle|}\int_{SO(3)}\,dR\int_{\R^3}\,du\, E\big(\Omega\cap (R\triangle+u)\big)-\text{error}$
\end{description}
which compares the energy of $\Omega$ with the energy of the reference set $\triangle$, averaged over rotations and translations of $\triangle$ inside $\Omega$, see Figure \ref{fig:A5}. This type of property was first remarked and used by Conlon, Lieb and Yau \cite{CLY1,CLY2}, for quantum systems interacting with the Yukawa potential and $\triangle$ being a cube. For Coulomb interactions, it was proved by Graf and Schenker \cite{GS,G} in which case $\triangle$ is chosen to be a tetrahedron. Fefferman also used a proof by means of a lower bound in his study of the crystal case in \cite{F}. This will be explained in details in \cite{2}.

It will be shown in Theorem \ref{limit_specific} that assumptions \textbf{(A1)--(A5)} are sufficient to obtain the existence of the limit
$$\lim_{n\to\ii}\frac{E(R_n\ell_n\triangle+u_n)}{\ell_n^3|\triangle|}=\bar e$$
for all sequences $\ell_n\to\ii$, $\{(R_n,u_n)\}\subset SO(3)\times\R^3$. 

\begin{figure}[h]
\centering
\includegraphics[width=8cm]{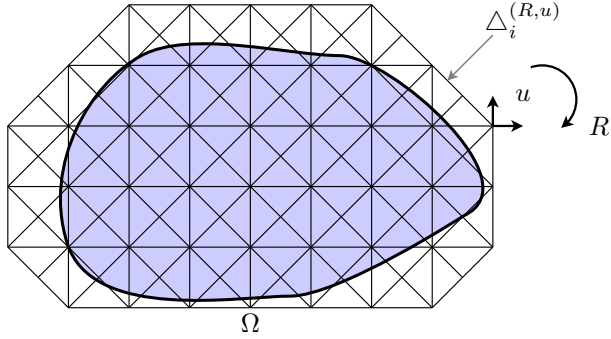}
\caption{Assumption \textbf{(A6)}: the energy in the big set $\Omega$ can be decomposed into energies in small sets $\triangle^{(R,u)}_i$ of the tiling plus the interaction between them and the difference between entropies.}
\label{fig:A6}
\end{figure}

In order to prove the existence of the limit for any sequence $\{\Omega_n\}$, we need the additional assumption that $\triangle$ is a polyhedron which can be used to build a tiling of $\R^3$ and the following new condition.
For all $\Omega\in\cR$, we assume that there exist numbers $I^{(R,u)}_{jk}$ and a function $S:\cM\to\R$ such that the energy can be decomposed in the form\footnote{In the text we indeed introduce two assumptions \textbf{(A6.1)} and \textbf{(A6.2)} corresponding to, respectively, a lower and an upper bound on the error term.}
\begin{description}
\item[(A6)] $\displaystyle E(\Omega)= \sum_iE\bigg(\Omega\cap\triangle_i^{(R,u)}\bigg)+\frac12\sum_{j\neq k}I^{(R,u)}_{jk}-S(\Omega)$\\ 
\hspace*{7cm}$\displaystyle+\sum_iS\bigg(\Omega\cap\triangle_i^{(R,u)}\bigg)+\text{error}$
\end{description}
where $\{\triangle_i^{(R,u)}\}$ are all the elements of the tiling, rotated by $R\in SO(3)$ and translated by $u\in\R^3$, see Figure \ref{fig:A6}.
The number $I_{jk}^{(R,u)}$ is interpreted as the interaction energy between the sets $\triangle_j^{(R,u)}$ and $\triangle_k^{(R,u)}$. The numbers $I^{(R,u)}_{jk}$ must satisfy some properties which we do not mention in this introduction, for this interpretation to be meaningful. 

The assumption which corresponds to \textbf{(A5)} is that the total interaction energy is nonnegative when averaged over translations and rotations of the tiling (the numbering of the equations corresponds to that of Section \ref{section_general_domains}):
\begin{description}
\item[(A6.3)] $\displaystyle \int_{}\,du\,dR\sum_{j\neq k}I^{(R,u)}_{jk}\geq-\text{error}.$
\end{description}
Finally, the function $S(\Omega)$ appearing in \textbf{(A6)} is assumed to satisfy certain properties similar to the ones of entropies in statistical mechanics. In the applications \cite{2}, $S$ will indeed just be the usual quantum entropy.
The main assumptions\footnote{In the text only the difference $S(\Omega)-\sum_iS(\Omega\cap\triangle_i^{(R,u)})$ is considered. See the details in Section \ref{section_general_domains}.} on $S$ are its normalization
\begin{description}
\item[(A6.5)] $S(\emptyset)=0$
\end{description}
and the \emph{strong subadditivity property:}
\begin{description}
\item[(A6.6)] $S(\Omega_1\cup\Omega_2\cup\Omega_3)+S(\Omega_2)\leq S(\Omega_1\cup\Omega_2)+S(\Omega_2\cup\Omega_3)$
\end{description}
for all disjoint subsets $\Omega_1$, $\Omega_2$ and $\Omega_3$. 

Conjectured by Lanford and Robinson \cite{LR} the strong subadditivity (SSA) of the quantum mechanical entropy  was proved by Lieb and Ruskai in \cite{LR1,LR2}. Robinson and Ruelle noticed in \cite{RR} (see also Wehrl \cite{Wehrl}) that SSA was a necessary property in order to define the entropy per unit volume of translation-invariant states: they used it to control boundary terms of any arbitrary Van Hove sequence of domains.
To our knowledge, SSA was never used in the way we do it to prove the thermodynamic limit for interacting systems.

\bigskip

The paper is organized as follows. In Section \ref{sec:general_theory} we state our main results, starting from the case of a special sequence $\Omega_n=R_n\ell_n\triangle+u_n$ for which we need less assumptions than for the general case. Proofs are given in Section \ref{sec:proofs_general}. Let us mention that the results of this paper and of \cite{2} have been summarized in \cite{proc}.

\bigskip

\noindent\textbf{Acknowledgment.} M.L. acknowledges support from the ANR project ``ACCQUAREL'' of the French ministry of research.

\section{General Theory of Thermodynamic Limit}\label{sec:general_theory}
\subsection{Definitions}
\subsubsection*{Regularity property of sets in $\R^3$}
Let us first define the class of sets for which we shall be able to prove the thermodynamic limit. In the whole paper, we denote by $|A|$ the Lebesgue measure of any measurable set $A\subseteq\R^3$, and by $\#B$ the cardinality of any finite set $B$. 
Let
$$\cM:=\left\{\Omega\subset\R^3\ |\ \Omega\ \text{is open and bounded}\right\}.$$
We shall restrict ourselves to the three dimensional case for simplicity but all the results of this first paper can easily be generalized to the $N$ dimensional case.
We shall need the following
\begin{definition}[Sets with regular boundary]\label{def_regular} Let $\eta:[0,t_0)\mapsto[0,\ii)$ be a real function\footnote{In the whole paper, we use the convention that $\eta$ is given with its domain of definition, i.e. with $t_0>0$. } with $\lim_{t\to0}\eta(t)=0$.
We say that $\Omega\in\cM$ has \emph{an $\eta$-regular boundary in the sense of Fisher} if
$$\forall t\in[0,t_0),\qquad \left|\left\{x\in\R^3\ |\ \textnormal{d}(x,\partial\Omega)\leq |\Omega|^{1/3}t\right\}\right|\leq |\Omega|\,\eta(t),$$
where $\partial\Omega=\overline{\Omega}\setminus{\Omega}$ is the boundary of ${\Omega}$.

We denote by $\cR_{\eta}\subset\cM$ the set of all subsets $\Omega\in\cM$ having an $\eta$-regular boundary in the sense of Fisher.
\end{definition}

The definition of $\eta$-regular boundary was introduced by Fisher \cite[page 394]{Fisher} (see also \cite[page 351]{LL}) and it is invariant by scaling.
In the following, we shall not consider any function $\eta$, but for simplicity only those belonging to the following class
$$\cE:=\left\{\eta:[0,c)\to[0,\ii)\ \big|\ \eta(t)=a\,t^b,\ a>0,\ c>0,\ b\in(0,1]\right\}.$$
Most of our results could be easily generalized to a more general class of functions $\eta$.
But the class of sets having Fisher's boundary property for some $\eta\in\cE$ already contains very ``perturbed" sets for which the area of the boundary can be of order greater than $|\Omega|^{2/3}$, see the footnote of \cite[page 394]{Fisher}.

Notice the following simple result, whose proof is given in Section \ref{sec_proof_convex}.
\begin{lemma}\label{regularity_convex}
Any open and bounded convex subset of $\R^3$ belongs to $\cR_\eta$ for all $\eta(t)=at$, $t\in[0,t_0)$, with $a$ large enough and $t_0$ small enough.
\end{lemma}

\subsubsection*{The sliding group}
We consider the group consisting of orientation preserving isometries of $\R^3$, i.e. the semidirect
product $G=\R^3\rtimes SO(3)$. This subgroup $G$ of the Euclidian group (it is sometimes called the \emph{Special Euclidian group}) acts on $\R^3$ as follows: $(u,R)\cdot x=Rx+u$. Let $d\lambda$ denote the Haar measure of $G$. Then
\begin{equation}
\int_G f(gy)d\lambda(g)=\int_{\R^3} f(x) dx
\label{prop_sliding_group}
\end{equation}
for any $f\in L^1(\R^3)$ and any $y\in\R^3$. 

Of course, $G$ also acts on $\cM$ and it stabilizes $\cR_{\eta}\subset\cM$ for any $\eta\in\cE$.

\subsection{Thermodynamic limit for a special sequence of sets}
Let us consider a function $E:\cM\to\R$. Our main goal is to prove the existence of a thermodynamic limit under suitable assumptions on $E$. This means that there exists a real number $\bar e$ such that 
$$\lim_{n\to\ii}\frac{E(\Omega_n)}{|\Omega_n|}=\bar e$$ 
for any `regular' sequence of sets $\{\Omega_n\}_{n\geq1}$ with $|\Omega_n|\to\ii$ as $n\to\ii$. 

The regularity of sets will be measured differently depending on the applications.
 For the general theory, let us consider some fixed function $\eta\in\cE$ defined on $[0,c)$, and an abstract class of sets $\cR\subseteq\cM$. Here we only assume that $\cR$ is invariant under translations, rotations and scaling, and that it contains sets having at least Fisher's $\eta$-regularity boundary property:
\begin{description}
\item[(P1)] One has $\cR\subseteq \cR_\eta$. If $\Omega\in\cR$, then $g\ell\Omega\in\cR$ for any $\ell\geq1$ and any $g\in G$. Moreover, $\emptyset\in\cR$.
\end{description}
For the crystal case, $\cR$ will consist of the sets having the cone property and a regular boundary, see \cite{2}. In some other cases, we shall be able to take $\cR=\cR_\eta$ for some fixed $\eta\in\cE$.
We also define the \emph{regularized volume} of any $\Omega\in\cM$ as follows
\begin{equation}
|\Omega|_{\rm r}:=\inf\{|\tilde\Omega|,\ \tilde\Omega\in\cR,\ \tilde\Omega\supseteq\Omega\},
\label{def_regularized_volume} 
\end{equation}
i.e. $|\Omega|_{\rm r}$ is the smallest volume of all regular sets $\tilde\Omega\in\cR$ containing $\Omega$. Notice the obvious inequality $|\Omega|_{\rm r}\geq |\Omega|$.

Also we shall need another class $\cR'$ of subsets of $\R^3$. We only assume at this stage that there exists some $\eta'\in\cE$ defined on $[0,c')$ with $\eta\leq\eta'$ on $[0,\min(c,c'))$ such that
\begin{description}
\item[(P2)] $\cR\subseteq\cR'\subseteq\cR_{\eta'}$.
\end{description}
The role of $\cR'$ will appear more clearly in the next section. For this section, the reader can simply take $\cR'=\cR$.

In this section, we start with the problem to prove the existence of the thermodynamic limit for specific sequences made of a reference set $\triangle\in\cR$, i.e. $\Omega_n=g_n \ell_n \triangle$ with $\{g_n\}\subseteq G$ and $\ell_n\to\ii$. We require adequate properties on the function $E$, which will need to be strengthened later to obtain the limit for general sequences.

For the moment, we only assume that the reference set $\triangle$ satisfies the following property:
\begin{description}
\item[(P3)] $\triangle\in\cR$ is a convex set such that $0\in\triangle$.
\end{description}
Thus $(1-\lambda)\triangle\subset\triangle$ for any $0<\lambda<1$. Later, we shall need to assume that $\triangle$ can be used to define a tiling of $\R^3$, but this is not necessary at this point. 

For the function $E$, we assume that there exists a subset $\cM_5\subseteq\cM$, a function $\hat{e}:\cR\to\R$, two positive constants $\delta,\kappa>0$ and a function $\alpha$ with $\lim_{\ell\to\ii}\alpha(\ell)=0$ such that the following hold for $\ell\geq1$:
\begin{description}
\item[(A1)] {\it (Normalization).} $E(\emptyset)=0$.
\item[(A2)] {\it (Stability).} For all $\Omega\in\cM$, we have
$$E(\Omega)\geq -\kappa|\Omega|.$$
\item[(A3)] {\it (Translation Invariance in Average).} For all $\Omega\in\cR$,
\begin{equation}
\left||B(0,L)|^{-1}\int_{B(0,L)}\frac{E(\Omega+u)}{|\Omega|}\,du-\hat{e}(\Omega)\right|\leq \alpha(L),
\label{weak_periodicity}
\end{equation}
where $B(0,L)$ denotes the ball of center $0$ and radius $L$.
\item[(A4)] {\it (Continuity).} For all $\Omega\in\cR$ and $\Omega'\in\cR'$ with $\Omega'\subseteq\Omega$ and $\text{d}(\partial\Omega,\partial\Omega')>\delta$, we have 
$$E(\Omega)\leq E(\Omega')+\kappa|\Omega\setminus\Omega'|+|\Omega|\alpha(|\Omega|).$$
\item[(A5)] {\it (Subaverage Property).} For all $\Omega\in\cM_5$, we have
\begin{equation}
E(\Omega)\geq \frac{1-\alpha(\ell)}{|\ell\triangle|}\int_{G}E\big(\Omega\cap g\cdot(\ell\triangle)\big)\,d\lambda(g)-|\Omega|_{\rm r}\,\alpha(\ell).
\label{sliding_equation}
\end{equation}
\end{description}

\begin{remark}\rm
 Notice that \textbf{(A1)}, \textbf{(A2)} and \textbf{(A4)} imply that there exists a constant $\kappa'>0$ such that 
\begin{equation}
-\kappa|\Omega|\leq E(\Omega)\leq \kappa'|\Omega|
\label{unif_bd}
\end{equation}
for all $\Omega\in\cR$ and for $|\Omega|$ large enough (take $\Omega'=\emptyset$ in \textbf{(A4)}). However the upper bound need not be true\footnote{In the crystal case studied in \cite{2}, the upper bound \emph{does not hold} for all $\Omega\in\cM$. For the other two examples of \cite{2} the upper bound is \emph{true}, more precisely one has $E(\Omega)\leq0$ for any $\Omega$ in $\cM$.} for all $\Omega$ in $\cM$. 
\end{remark}

\begin{remark}\rm
 Assumption \textbf{(A3)} is a generalization of the periodicity requirement. Examples of functions $E$ satisfying \textbf{(A3)} contain
\begin{enumerate}
\item functions satisfying \eqref{unif_bd} and which are $\Lambda$-periodic for some discrete subgroup $\Lambda\subseteq\R^3$ with compact fundamental domain $\R^3/\Lambda$, hence in particular fully translation-invariant systems, $\forall u\in\R^3,\ E(\Omega+u)=E(\Omega)$;
\item finite sums of periodic functions as above;
\item functions which are almost $\Lambda$-periodic, uniformly with respect to $\Omega$.
\end{enumerate}
In the proof of our results, we actually do not need \textbf{(A3)} for all $\Omega\in\cR$, but only for $\Omega=g\ell\triangle$ with $g\in G$ and $\ell$ large enough.
\end{remark}

\begin{remark}\rm
 The most important assumption is of course \textbf{(A5)} which allows to bound from below the energy in a large set $\Omega$ by an average of energies of the (smaller) scaled reference set $\ell\triangle$, as explained in introduction. We introduced the set $\cM_5$ for which property \textbf{(A5)} is true. In this section we shall simply take $\cM_5=\cM_\triangle$ with
\begin{equation}
\cM_\triangle:=\left\{g\ell\triangle,\quad g\in G,\ \ell\geq\ell_0\right\}
\label{choice_A_5}
\end{equation}
for some fixed $\ell_0\geq1$, but in the limit for general sets in the next section, we shall need \textbf{(A5)} for all domains $\Omega$ (i.e. we will choose $\cM_5=\cM$). Notice in all cases the error is assumed to depend on $|\Omega|_{\rm r}$, that is to say on the volume of the smallest regular set containing $\Omega$.
\end{remark}

\begin{remark}\rm
 Property  \textbf{(A4)} is also very important. It essentially says that a small decrease of $\Omega$ will not decrease too much the energy. A similar property was used and proved in the crystal case by Fefferman \cite[Lemma 2]{F}. In the proof of the existence of the thermodynamic limit for general domains, we shall need this property for domains $\Omega'$ which are a bit less regular than sets in $\cR$ (see Section \ref{section_general_domains}), explaining the introduction of the class $\cR'$. For the result of this section, we could simply take $\cR'=\cR$ but later on we shall need to cover the case $\cR'\varsupsetneq\cR$.
\end{remark}

\begin{remark}\rm
 Notice that the function $g\mapsto E\big(\Omega\cap g\cdot(\ell\triangle)\big)$ has compact support by \textbf{(A1)} and is bounded below by \textbf{(A2)}:
$$E\big(\Omega\cap g\cdot(\ell\triangle)\big)\geq -\kappa |\Omega|.$$
Therefore \textbf{(A5)} in particular implies that $g\mapsto E\big(\Omega\cap g\cdot(\ell\triangle)\big)$ belongs to $L^1(G,d\lambda)$ for any $\Omega\in\cM_5$.
\end{remark}

\begin{remark}\rm
 Assume that $E$ is a local functional of the form
\begin{equation}
E(\Omega)=\int_{\Omega}\chi(x)\,dx
\label{form_fn_local}
\end{equation}
for some fixed function $\chi\in L^\ii(\R^3,\R)$ satisfying
\begin{equation}
\lim_{L\to\ii}\left||B(0,L)|^{-1}\int_{B(0,L)}\chi(x+u)\,du-\chi_\ii(x)\right|=0
\label{average_pty}
\end{equation}
for some function $\chi_\ii:\R\mapsto\R$, uniformly with respect to $x$ (it can be proved that $\chi_\ii$ is necessarily constant, see Lemma \ref{translation_invariance_f} below). See \cite{BLL} for comments on the property \eqref{average_pty}.
Then \textbf{(A1)}--\textbf{(A5)} are satisfied for any $\Omega\in\cM$ and any reference set $\triangle$. In particular \textbf{(A5)} is always an equality, as
\begin{equation}
\frac{1}{|A|}\int_{G}E\big(\Omega\cap g\cdot A\big)\,d\lambda(g) = \frac{1}{|A|}\int_{\R^3}dx\int_Gd\lambda(g)\1_\Omega(x)\1_{A}(g^{-1}x)\chi(x) = E(\Omega)
\label{calcul_local_fn}
\end{equation}
for any set $A\in\cM$, and where we have used that 
$$\int_Gd\lambda(g)\1_{A}(g^{-1}x)=|A|,$$
see \eqref{prop_sliding_group}. For the volume $E(\Omega)=|\Omega|$, $\chi\equiv 1$, this in particular gives
\begin{equation}
|\Omega|=\frac{1}{|\ell\triangle|}\int_Gd\lambda(g)\big|\Omega\cap g\cdot(\ell\triangle)\big|.
\label{eq_volume}
\end{equation}
If a function $E$ takes the form \eqref{form_fn_local} for some $\chi\in L^\ii(\R^3,\R)$ satisfying \eqref{average_pty}, then the thermodynamic limit is easily proved to be the constant $\chi_\ii$.
\end{remark}

\medskip

We are now able to state our
\begin{thm}[Thermodynamic limit for the reference set $\triangle$]\label{limit_specific} We assume that $\cR$, $\cR'$ and $\triangle$ satisfy \textnormal{\textbf{(P1)}--\textbf{(P3)}} and that the function $E$ satisfies \textnormal{\textbf{(A1)}--\textbf{(A5)}} with $\cM_5=\cM_\triangle$ defined in \eqref{choice_A_5}. We denote 
$$e_\ell(g)=\frac{E\big(g\ell\triangle\big)}{|\ell\triangle|}$$ 
for $g\in G=\R^3\rtimes SO(3)$ and $\ell\geq1$.

Then there exists a real constant $\bar e$ such that $e_\ell$ converges uniformly to $\bar e$ on $G$ as $\ell\to\ii$. In particular, for all sequences $\{\ell_n\}_{n\geq1}\subseteq[1,\ii)$ with $\ell_n\to\ii$ and $\{g_n\}_{n\geq1}\subseteq G$,
$$\lim_{n\to\ii}\frac{E\big(g_n\ell_n\triangle\big)}{|\ell_n\triangle|}=\bar e.$$

Additionally, the limit $\bar e$ does not depend on the reference set $\triangle$: if the above properties hold true for another set $\triangle'$ and if $\cM_5=\cM_\triangle\cup\cM_{\triangle'}$, then ${E\big(g\ell\triangle'\big)}{|\ell\triangle'|^{-1}}$ converges to the same limit $\bar e$ as $\ell\to\ii$, uniformly on $G$.

\end{thm}

The proof of Theorem \ref{limit_specific} is given in Section \ref{proof_specific} below.

\begin{remark}\rm
We notice that Theorem \ref{limit_specific} holds true if we replace \textbf{(A2)} by the weaker assumption involving the regularized volume
\begin{description}
\item[(A2')] {\it (Weak Stability).} For any $\Omega\in\cM$, then
$$E(\Omega)\geq -\kappa|\Omega|_{\rm r}.$$
\end{description}
However \textbf{(A2)} will be needed in the next section where we tackle the case of a general sequence of domains. 
\end{remark}

\subsection{Thermodynamic limit for general sets}\label{section_general_domains}
In Theorem \ref{limit_specific}, we have stated the existence of the thermodynamic limit for domains $\triangle$ satisfying certain assumptions. However in the applications it is not clear for which domain $\triangle$, property \textbf{(A5)} will hold true. At present, for Coulomb systems an inequality of the form \textbf{(A5)} is only known for simplices \cite{GS,2}. Hence, in order to prove the existence of the thermodynamic limit for any (regular) sequence of sets, we need to add some assumptions. The idea will be to decompose any domain as a union of (translated and rotated) reference sets $\triangle$. For this reason we will now assume that $\triangle$ can be used to build a tiling of the whole space, and we will require some localization properties on the energy $E$. 
First, we introduce the following
\begin{definition}[Tilings of $\R^3$] Let $\Gamma$ be a discrete subgroup of $G$ with compact quotient $G/\Gamma$. We say that the open set $\triangle\subset\R^3$ \emph{defines a $\Gamma$-tiling of $\R^3$} if 
\begin{itemize}
\item $\mu \triangle\cap\mu'\triangle=\emptyset$ if
  $\mu,\mu'\in \Gamma$ and $\mu\ne\mu'$;
\item $\bigcup_{\mu\in\Gamma}\overline{\mu\triangle}=\R^3$.
\end{itemize}
\end{definition}

In the following, we assume additionally to \textbf{(P1)}--\textbf{(P3)} that 
\begin{description}
\item[(P4)] $\triangle$ is a polyhedron which defines a $\Gamma$-tiling for some discrete subgroup $\Gamma$ of $G$ with $G/\Gamma$ compact.
\end{description}
In the Coulomb case \cite{GS,G}, we shall essentially (up to a translation) take $\Gamma=\Z^3\rtimes O$, where $O$ is the
group of order 24 consisting of the pure rotations of the octahedral (or hexahedral) group, i.e. the symmetry group of the cube. The set $\triangle$ is the simplex which makes up a 24th of the cube.

\begin{remark}\rm
Notice that there is no loss of generality in assuming that the convex set $\triangle$ is a polyhedron: it can easily be proved by means of the finite-dimensional Hahn-Banach Theorem that any convex set defining a $\Gamma$-tiling is necessarily a polyhedron.
\end{remark}

\begin{figure}[h]
\centering
\includegraphics[width=7cm]{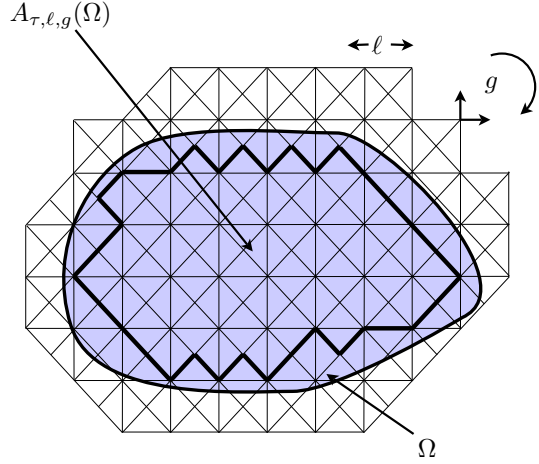}
\caption{The inner approximation $A_{\tau,\ell,g}(\Omega)$ of a set $\Omega$ (with $\tau=0$ here).}
\label{fig:inner_approx}
\end{figure}

Given $\triangle$ which defines a $\Gamma$-tiling, any regular set $\Omega\in\cR$ can be approximated by unions of elements of the (dilated and rotated) tiling, see Figure \ref{fig:inner_approx}. The following proposition tells us that the so-approximated set has a boundary which stays regular, although it might be a bit less regular than that of $\Omega$. It will be very useful in the proof of our results.

\begin{prop}[Regularity of the boundary of the inner approximation]\label{reg_inner_approx} Let $\delta,\tau_0>0$, $\Omega\in\cM$ and introduce for any $g\in G$, $\tau,\ell>0$
$$\triangle_{\tau,\ell,g}(\mu)=\ell g\mu(1+\tau)\triangle.$$
If 
\begin{equation}
A_{\tau,\ell,g}(\Omega)=\bigcup_{\substack{\mu\in\Gamma,\\ \triangle_{\tau,\ell,g}(\mu)\subset\Omega,\\ \text{d}(\partial\triangle_{\tau,\ell,g}(\mu),\partial\Omega)>\delta}}\triangle_{\tau,\ell,g}(\mu)\subseteq\Omega,
\label{def_approx_tiling_general}
\end{equation}
then there exists three constants $\ell_0>0$, $\ell_1>0$ and $m\geq1$ such that 
$$\left\{A_{\tau,\ell,g}(\Omega)\ |\ \Omega\in\cR_\eta,\ g\in G,\  0\leq \tau\leq \tau_0,\ \ell_0\leq\ell\leq|\Omega|^{1/3}\ell_1\right\}\subseteq \cR_{\tilde{\eta}},$$
with $\tilde{\eta}(t)=m\eta(t),\ t\in[0,c/m)$ where we recall that the domain of definition of $\eta$ is $[0,c)$.
\end{prop}

\begin{remark}\rm
Since $\delta$ will be fixed in accordance with \textbf{(A4)}, we do not explicitely emphasize that $A_{\tau,\ell,g}(\Omega)$ depends on $\delta$.
\end{remark}

Proposition \ref{reg_inner_approx} is shown in Section \ref{proof_reg_inner_approx}.
In the proof of our result, we shall need to know that the energy $E(\Omega)$ does not vary too much when $\Omega$ is replaced by its inner approximation $\Omega'=A_{\tau,\ell,g}(\Omega)$. This will be true provided we can apply \textbf{(A4)}, leading to the natural assumption that $A_{\tau,\ell,g}(\Omega)$ belongs to $\cR'$ for any $\Omega\in\cR$ (the constant $\delta$ is chosen like in \textbf{(A4)}):
\begin{description}
\item[(P5)] For any $\tau_0>0$, there exists two constants $\ell_0>0$, $\ell_1>0$ such that
$$\left\{A_{\tau,\ell,g}(\Omega)\ |\ \Omega\in\cR,\ g\in G,\  0\leq \tau\leq \tau_0,\ \ell_0\leq\ell\leq|\Omega|^{1/3}\ell_1\right\}\subseteq \cR'.$$
\end{description}

We need to add some assumptions on the energy $E$ to prove the existence of the thermodynamic limit for general sets. We assume that there exists a class of domains $\cM_6\subset\cM$ such that the following holds
\begin{description}
\item[(A6)] \textit{(Local Energy Decomposition).} For all $\Omega\in\cM_6\subset\cM$  and all $\ell\geq1$ there exist maps
$$E^\Omega_\ell:G\to\R,$$
$$ I^\Omega_\ell:\{(g,g')\in G\times G\ | \ g^{-1}g'\in\Gamma\setminus\{0\}\}\to\R$$
$$s^\Omega_\ell:G\times\{\cP\subseteq\Gamma\}\to\R$$
such that the following conditions hold for all $g\in G$:
\begin{description}
\item[(A6.1)] {\it (Lower bound).} 
\begin{equation}
E(\Omega)\geq 
\sum_{\mu\in\Gamma}E^\Omega_\ell(g\mu)+
\frac{1}{2}\sum_{\mu,\nu\in\Gamma\atop\mu\ne\nu}
I^\Omega_\ell(g\mu ,g\nu) -s^\Omega_\ell(g,\Gamma)-|\Omega|\alpha(\ell).
\label{lower_bound_Omega}
\end{equation}
\item[(A6.2)] {\it (Upper bound).} For all $\cP\subseteq\Gamma$,
\begin{multline*}
E\left(\Omega\,\cap\bigcup_{\mu\in\cP}\ell g\mu (1+\tau(\ell))\triangle\right)\leq 
\sum_{\mu\in\cP}E^\Omega_\ell(g\mu)+
\frac{1}{2}\sum_{\mu,\nu\in\cP\atop \mu\ne\nu}
I^\Omega_\ell(g\mu ,g\nu)\\
 -s^\Omega_\ell(g,\cP)+\left|\Omega\,\cap\bigcup_{\mu\in\cP}\ell g\mu(1+\tau(\ell))\triangle\right|\alpha(\ell),
\end{multline*}
for some continuous function $\tau:(0,\ii)\to[0,\ii)$ which does not depend on $\Omega$ and satisfies $\lim_{\ell\to\ii}\tau(\ell)=0$.
\item[(A6.3)] {\it (Interaction Average).} 
$$\frac1{|\triangle|}\int_{G}d{\lambda}(g)
  \sum_{\mu\in\Gamma\atop\mu\ne 0}I^\Omega_\ell(g\mu,g)
\geq -|\Omega|\alpha(\ell).$$
\item[(A6.4)] {\it (Interaction Support).} $I_\ell^\Omega(g\mu,g\nu)=0$ if $\ell g\mu(1+\tau(\ell))\triangle\cap\Omega=\emptyset$ or if $\ell g\nu(1+\tau(\ell))\triangle\cap\Omega=\emptyset$.
\item[(A6.5)] {\it (Normalization of $s^\Omega_\ell$).} For any $\mu\in\Gamma$ and $g\in G$,
\begin{equation*}
s^\Omega_\ell(g,\{\mu\})=s^\Omega_\ell(g,\emptyset)=0. 
\label{normalization_A6_5}
\end{equation*}
\item[(A6.6)] {\it (Strong Subadditivity of $s^\Omega_\ell$).} For any disjoint subsets $\cP_1$, $\cP_2$, $\cP_3\subseteq\Gamma$ and any $g\in G$,
\begin{equation*}
s^\Omega_\ell(g,\cP_1\cup\cP_2\cup\cP_3)+s^\Omega_\ell(g,\cP_2)\leq s^\Omega_\ell(g,\cP_1\cup\cP_2)+s^\Omega_\ell(g,\cP_2\cup\cP_3).
\label{strong_subadd_A6_5} 
\end{equation*}
\end{description}
\end{description}

Let us make some comments. Assumption \textbf{(A6)} essentially means that our energy arises from a physical system containing at most two-body interactions. 
The reader has to imagine that $E^\Omega_\ell(g\mu)$ represents the total (free) energy in each  $\ell g\mu\triangle$, whereas $I_\ell^\Omega(g\mu ,g\nu)$ describes the interaction energy between $\ell g\mu\triangle$ and $\ell g\nu\triangle$. 
Our main result below can easily be extended to the case of a $k$-body interaction but we do not give further details as this would complicate \textbf{(A6)}.

Assumption \textbf{(A6.3)} means that averaging over translations and rotations of the tiling yields a global interaction energy which is essentially nonnegative. This is a specified averaging-type property which is more precise than our assumption \textbf{(A5)} above. Indeed, we shall prove that \textbf{(A6)} is stronger than \textbf{(A5)}. This is expressed in the following result, whose proof will be given in Section \ref{sec:proof_A5_implies_A6}.
\begin{prop}\label{A6_implies_A5}
Let $\cR$, $\cR'$ and $\triangle$ satisfy \textnormal{\textbf{(P1)}--\textbf{(P5)}}, and $E$ satisfying \textnormal{\textbf{(A1)}--\textbf{(A4)}}. If $E$ satisfies \textnormal{\textbf{(A6)}} for all $\Omega\in\cM_6$ where $\cM_6\subseteq\cM$, then $E$ also satifies the averaging property \textnormal{\textbf{(A5)}} for all $\Omega\in\cM_6$. This means we can assume that $\cM_6\subseteq\cM_5$
\end{prop}

In our study, we shall need \textbf{(A5)} for all domains $\Omega\in\cM$ whereas we only use \textbf{(A6)} for $\Omega\in\cM_\triangle$ which was defined in \eqref{choice_A_5}. If \textbf{(A6)} is true for all domains $\Omega\in\cM$, we do not need assumption \textbf{(A5)} by Proposition \ref{A6_implies_A5}.

In the applications, the function $s^\Omega_\ell$ will be related to the entropy of the system. Formally, it will be the difference between the entropy in the big domain and the entropies of the small sets:
$$s^\Omega_\ell(g,\cP)=S\left( \Omega\,\cap\bigcup_{\mu\in\cP}\ell g\mu \triangle\right) - \sum_{\mu\in\cP} S\left( \Omega\,\cap\ell g\mu \triangle\right).$$

 For the original tiling $\{\ell g\mu\triangle\}_{\mu\in\Gamma}$, it might be difficult to prove that the energy behaves as stated, in particular because of the considered boundary conditions and localization issues. For this reason, one may need that each of the sets used for the tiling slightly overlap, which is the role of the correction $\tau(\ell)$. Notice that since $0\in\triangle$, then $\triangle\subset(1+\tau(\ell))\triangle$.

\begin{remark}\rm 
For the sake of simplicity, we have used the same function $\alpha$ as in the previous section. We recall that $\lim_{\ell\to\ii}\alpha(\ell)=0$. Symetrizing the above properties if necessary, we can always assume that $I^\Omega_\ell(\cdot,\cdot)$ is symmetric: 
$$\forall g\in G,\ \forall\mu,\nu\in\Gamma,\ \mu\neq\nu,\quad  I_\ell^\Omega(g\mu,g\nu)=I_\ell^\Omega(g\nu,g\mu).$$
Note that it would be natural to assume in addition to \textbf{(A6.4)} that $E(g\mu)=0$ when $\ell g\mu(1+\tau(\ell))\triangle\cap\Omega=\emptyset$ and that $s^\Omega_\ell(g,\cP)=0$ when $\ell g\mu(1+\tau(\ell))\triangle\cap\Omega=\emptyset$ for any $\mu\in\cP$, but we actually do not need it in the proof.
\end{remark}

\begin{remark}\rm 
The stability condition \textbf{(A2)} is used in the proof to control some complicated boundary terms. Our proof does not work under the weaker assumption \textbf{(A2')} introduced in the previous section.
\end{remark}

\begin{remark}\rm 
Taking $\cP_2=\emptyset$ in \textbf{(A6.6)} and using \textbf{(A6.5)}, we deduce that $s^\Omega_\ell$ is subadditive:
\begin{equation}
 s^\Omega_\ell(g,\cP_1\cup\cP_3)\leq s^\Omega_\ell(g,\cP_1)+s^\Omega_\ell(g,\cP_3).
\label{subadd_entropy_general}
\end{equation}
Then, using that $s^\Omega_\ell(g,\{\mu\})=0$ for any $\mu\in\Gamma$ by \textbf{(A6.5)}, we also deduce by induction that $s$ is monotone
\begin{equation}
\cP_1\subseteq\cP_2\subseteq\Gamma \Longrightarrow s^\Omega_\ell(g,\cP_2)\leq s^\Omega_\ell(g,\cP_1),
\label{s_monotone}
\end{equation}
hence that
\begin{equation}
\forall \cP\subseteq\Gamma,\qquad s^\Omega_\ell(g,\cP)\leq0
\label{nonpositivity}
\end{equation}
for any $g\in G$.
\end{remark}

\begin{remark}\rm 
In the proof we do not use the full strong subadditivity \textbf{(A6.6)}, but rather the weaker property
\begin{equation}
 s_\ell^\Omega(g,\cP)\leq \frac{1}{\#\cP}\sum_{\mu\neq\nu\in\cP}s_\ell^\Omega(g,\{\mu,\nu\})
\label{estim_SSA_weak}
\end{equation}
which is satisfied for any strongly subadditive function satisfying \textbf{(A6.5)}, as proved in Lemma \ref{lemmaT}. However the generalization to a $k$-body interaction would require an estimate different from \eqref{estim_SSA_weak} and for this reason we prefer to keep the more general assumption \textbf{(A6.6)}. Also it is well-known that the strong subadditivity of the entropy is a useful tool in the study of thermodynamic limits \cite{RR,Wehrl}.
\end{remark}

\medskip

We are now able to state our main
\begin{thm}[Thermodynamic limit for general domains]\label{limit_general} We assume that $\cR$, $\cR'$ and $\triangle$ satisfy \textnormal{\textbf{(P1)}--\textbf{(P5)}} and that the function $E$ satisfies \textnormal{\textbf{(A1)}--\textbf{(A6)}} with $\cM_5=\cM$ and $\cM_6=\cM_\triangle$.

Then, for any sequence $\{\Omega_n\}_{n\geq1}\subseteq\cR$ such that  
${{\rm diam}(\Omega_n)} |\Omega_n|^{-1/3}$ is bounded and $|\Omega_n|\to\ii$, we have
$$\lim_{n\to\ii}\frac{E(\Omega_n)}{|\Omega_n|}=\bar e,$$
where $\bar e$ is the limit obtained in Theorem \ref{limit_specific}.
\end{thm}

We recall that $\cM_\triangle$ was defined in \eqref{choice_A_5}. The rest of the paper is devoted to the proof of our results. Applications are given in the second part \cite{2}.

\begin{remark}\rm
As we have $\cR\subset\cR_\eta$, we know from \cite[Appendix A p.~385]{LL} and \cite[Lemma 1]{Fisher} that if each set $\Omega_n$ is connected, then automatically ${\rm diam}(\Omega_n)|\Omega_n|^{-1/3}\leq C$ for all $n$. 
\end{remark}

\section{Proofs}\label{sec:proofs_general}
\subsection{Proof of Theorem \ref{limit_specific}: limit for the reference set}\label{proof_specific}

Notice that applying \eqref{unif_bd} to $g\ell\triangle$ which is in $\cR$ by \textbf{(P1)} and \textbf{(P3)}, we infer that $e_\ell$ is uniformly bounded on $G$.

\medskip

\noindent{\textbf{Step 1.} \textit{Translation invariance of $\hat{e}$ defined in \textnormal{\textbf{(A3)}}.}} 
We start by proving the following
\begin{lemma}\label{translation_invariance_f} For all $\Omega\in\cR$ and all $u\in\R^3$, one has $\hat{e}(\Omega+u)=\hat{e}(\Omega)$.
\end{lemma}
\begin{proof}
Let $\Omega\in\cR$ and $u\in\R^3$. Then, for $L$ large enough,
\begin{multline*}
\left||B(0,L)|^{-1}\int_{B(0,L)}\frac{E(\Omega+v)}{|\Omega|}\,dv-|B(0,L)|^{-1}\int_{B(u,L)}\frac{E(\Omega+v)}{|\Omega|}\,dv\right|\\
\leq \max(\kappa,\kappa')\frac{|B(0,L)\setminus B(u,L)|+|B(u,L)\setminus B(0,L)|}{|B(0,L)|}
\end{multline*}
by \eqref{unif_bd} and since $\Omega+w\in\cR$ for any $w\in\R^3$ by \textbf{(P1)}. The result is obtained by taking the limit $L\to\ii$.
\end{proof}

\medskip

\noindent{\textbf{Step 2.} \textit{Lower bound.}} We now prove the important 
\begin{lemma}[Bound from below]\label{lemma_estim_below_average}There exist a decreasing function $\alpha_1$ tending to zero at infinity and a constant $\kappa_1\geq1$ such that, for all $\tilde\eta\in\cE$ (defined on $[0,c')$), all $\Omega\in\cR_{\tilde\eta}\cap\cM_5$ and all $\ell\geq1$,
\begin{equation}
\frac{E(\Omega)}{|\Omega|} \geq e_\ell^{\rm av}-\frac{|\Omega|_{\rm r}}{|\Omega|}\alpha_1(\ell) -\kappa_1\tilde\eta\left(\frac{\ell}{|\Omega|^{1/3}}\right)
\label{estim_below_average}
\end{equation}
provided $\ell|\Omega|^{-1/3}< c'/\kappa_1$, and with
\begin{equation}
e_\ell^{\rm av}:=\int_{SO(3)}\hat{e}(R\ell\triangle)dR.
\label{def_e_av}
\end{equation}
\end{lemma}

\begin{proof}
Let $\Omega\in\cR_{\tilde\eta}\cap\cM_5$ and $\ell\geq1$. We introduce the set of `interior points'
$$I_\ell=\big\{u\in\R^3\ |\ (R\ell\triangle+u)\subseteq\Omega,\ \forall R\in SO(3)\big\}.$$
and of `boundary points'
$$B_\ell=\big\{u\in\R^3\ |\ \exists R\in SO(3),\ (R\ell\triangle+u)\cap\Omega\neq\emptyset\big\}\setminus I_\ell.$$
Notice that since $0\in\triangle$, then $I_\ell\subseteq\Omega$. By \textbf{(A5)}, we obtain
\begin{eqnarray}
E(\Omega) & \geq & (1-\alpha(\ell))\int_{I_\ell\times SO(3)}d\lambda(g) \frac{E\left(g\ell\triangle\right)}{|\ell\triangle|}-|\Omega|_{\rm r}\alpha(\ell)\nonumber\\
 &  & \qquad +(1-\alpha(\ell))\int_{B_\ell\times SO(3)}d\lambda(g) \frac{E\left(\Omega\cap(g\ell\triangle)\right)}{|\ell\triangle|}\\
 & \geq & (1-\alpha(\ell))\int_{SO(3)}dR\int_{I_\ell}du \frac{E\left(R\ell\triangle+u\right)}{|\ell\triangle|}-|\Omega|_{\rm r}\alpha(\ell)\nonumber\\
 & & \qquad -(1-\alpha(\ell))\kappa|B_\ell|.\label{decomposition_int_bdry}
\end{eqnarray}
where we have also used \textbf{(A2)}. 

We now use that \textbf{(A5)} is an equality for the local function 
$\chi:u\mapsto E\left(R\ell\triangle+u\right) |\ell\triangle|^{-1}$
 and with the ball $B(0,1):=A$ as reference set, as explained in \eqref{calcul_local_fn}. We obtain
\begin{eqnarray}
\int_{I_\ell}du\frac{E\left(R\ell\triangle+u\right)}{|\ell\triangle|} & = & \frac{1}{|B(0,{\ell})|}\int_Gd\lambda(g)\int_{I_\ell\cap gB(0,{\ell})}\frac{E\left(R\ell\triangle+u\right)}{|\ell\triangle|}du\nonumber\\
 & = & \frac{1}{|B(0,{\ell})|}\int_{\R^3}dx\int_{I_\ell\cap B(x,{\ell})}\frac{E\left(R\ell\triangle+u\right)}{|\ell\triangle|}du.\label{calcul_decomp_boule}
\end{eqnarray}
We introduce similarly to $I_\ell$ and $B_\ell$,
$$I'_\ell:=\big\{x\in\R^3\ |\ B(x,{\ell})\subseteq I_\ell\big\},$$
$$B'_\ell:=\big\{x\in\R^3\ |\ B(x,{\ell})\cap \Omega\neq\emptyset\big\}\setminus I'_\ell.$$
Notice that $I'_\ell\subseteq I_\ell\subseteq\Omega$ and $B'_\ell\subseteq B_{a\ell}$ for some constant $a$ depending only on $\triangle$. Then \eqref{calcul_decomp_boule} becomes by \textbf{(A2)}
$$\int_{I_\ell}du\frac{E\left(R\ell\triangle+u\right)}{|\ell\triangle|}\geq \int_{I'_\ell}dx\frac{1}{|B(0,{\ell})|}\int_{ B(x,{\ell})}\frac{E\left(R\ell\triangle+u\right)}{|\ell\triangle|}du-\kappa|B_{a\ell}|.$$
Using \textbf{(A3)} and Lemma \ref{translation_invariance_f}, we obtain that
$$\int_{I_\ell}du\frac{E\left(R\ell\triangle+u\right)}{|\ell\triangle|}\geq |I'_\ell|\hat{e}(R\ell\triangle)-|I'_\ell|\alpha({\ell})-\kappa|B_{a\ell}|.$$
Inserting in \eqref{decomposition_int_bdry}, we arrive at
$$E\left(\Omega\right)\geq (1-\alpha(\ell))|I'_\ell|e_\ell^{\rm av}-2|\Omega|_{\rm r}\alpha(\ell)-2\kappa|B_{a\ell}|.$$
Clearly 
\begin{equation}
|I'_\ell|\leq|\Omega|\leq |I'_\ell|+|B_{a\ell}|.
\label{estim_volume_I_ell}
\end{equation}
Using then that $|e_\ell^{\rm av}|\leq \max(\kappa,\kappa')$, we obtain that 
$$\frac{E\left(\Omega\right)}{|\Omega|}\geq e_\ell^{\rm av}-C\frac{|\Omega|_{\rm r}}{|\Omega|}\sup_{t\geq\ell}\{\alpha(t)\}-C\frac{|B_{a\ell}|}{|\Omega|}$$
for some uniform constant $C\geq1$.

Finally, we use that $\Omega\in\cR_{\tilde\eta}$ to estimate $|B_{a\ell}|$. It is clear that there is a constant $a'\geq1$ (depending only on $\triangle$) such that
$$B_{a\ell}\subseteq \left\{x\in\R^3\ |\ d(x,\partial\Omega)\leq a'\ell\right\},$$
thus, for $\kappa_1:=a'C$ and $\kappa_1\ell|\Omega|^{-1/3}<c'$ (recall $\tilde{\eta}$ is defined on $[0,c')$),
$$|B_{a\ell}|\leq |\Omega|\tilde\eta\left(\frac{a'\ell}{|\Omega|^{1/3}}\right)\leq |\Omega|a'\,\tilde\eta\left(\frac{\ell}{|\Omega|^{1/3}}\right)$$
which ends the proof of Lemma \ref{lemma_estim_below_average}.
\end{proof}

\medskip

\noindent{\textbf{Step 3.} \textit{Convergence of $e_\ell^{\rm av}$ and $\inf_G e_\ell$.}} Let us introduce
$$e_\ell^{\rm m}=\inf_{g\in G}e_\ell(g),$$
where we recall that $e_\ell(g)=E(g\ell\triangle)|\ell\triangle|^{-1}$.
Notice that by \textbf{(A3)} (and taking $L\to\ii$),
$$e_\ell^{\rm av}\geq e_\ell^{\rm m}.$$
By Lemma \ref{lemma_estim_below_average} (applied with $\tilde\eta=\eta$ and $\Omega=\ell'\triangle$), we have for any $\ell,\ell'\geq1$,
\begin{equation}
e_{\ell}^{\rm m}-\alpha_1(\ell)-\kappa_1\eta\left(\kappa_2\ell/\ell'\right)\\ \leq e_{\ell}^{\rm av}-\alpha_1(\ell)-\kappa_1\eta\left(\kappa_2\ell/\ell'\right)
 \leq  e_{\ell'}^{\rm m} \leq  e_{\ell'}^{\rm av}
\label{estim_av_m_encadrement}
\end{equation}
where $\kappa_2=|\triangle|^{-1/3}$ and provided $\kappa_1\kappa_2\ell/\ell'< c$.
Let $(\ell_n)$ and $(\ell'_n)$ be two sequences such that 
$$\lim_{n\to\ii}e_{\ell'_n}^{\rm m}=\liminf_{\ell'\to\ii}e_{\ell'}^{\rm m}\quad \text{and}\quad \lim_{n\to\ii}e_{\ell_n}^{\rm m}=\limsup_{\ell\to\ii}e_{\ell}^{\rm m}.$$
 Extracting subsequences, we can assume that $\ell_n/\ell'_n\to0$ as $n\to\ii$. By \eqref{estim_av_m_encadrement}, we deduce that $\limsup_{\ell\to\ii}e_{\ell}^{\rm m}\leq \liminf_{\ell\to\ii}e_{\ell}^{\rm m}$ and thus that  $\lim_{\ell\to\ii} e_\ell^{\rm m}:=\bar e$ exists. By the same argument and \eqref{estim_av_m_encadrement}, it coincides with $\lim_{\ell\to\ii}e_\ell^{\rm av}$.
Therefore, we have proved that there exists $\bar e$ such that
\begin{equation}
\lim_{\ell\to\ii}e_\ell^{\rm m}=\lim_{\ell\to\ii}e_\ell^{\rm av}=\bar e.
\label{limit_av_min}
\end{equation}
The same argument applied to
$$\hat{e}^{\rm m}_\ell:=\inf_{R\in SO(3)}\hat{e}(R\ell\triangle)\leq e_\ell^{\rm av}$$
shows
\begin{equation}
\lim_{\ell\to\ii}\hat{e}_\ell^{\rm m}=\bar e.
\label{limit_av_min2}
\end{equation}

We end this step by proving that $\hat{e}_\ell(R):=\hat{e}(R\ell\triangle)$ converges to $\bar e$ in $L^1(SO(3),dR)$. Indeed, let us introduce $g_\ell:=\hat{e}_\ell-\bar e$ and write $g_\ell=g_\ell^+-g_\ell^-$ with $g_\ell^+,g_\ell^-\geq0$ and $g_\ell^+g_\ell^-=0$. Then, since $g_\ell^-=g_\ell\times\1_{\hat{e}_\ell^{\rm m}\leq \hat{e}_\ell(\cdot)\leq \bar e}$,
we infer
$$0\leq g_\ell^-\leq |\hat{e}_\ell^{\rm m}-\bar e|$$
which proves that $g_\ell^-$ converges uniformly to $0$ on $SO(3)$. Now,
$$\int_{SO(3)}g_\ell^+=\int_{SO(3)}g_\ell^- + (e_\ell^{\rm av}-\bar e)$$
which converges to 0 as $\ell\to\ii$ and shows
$$\lim_{\ell\to0}\norm{\hat{e}_\ell-\bar e}_{L^1(SO(3),dR)}=0.$$

\medskip

\noindent{\textbf{Step 4.} \textit{Uniform convergence of $e_\ell$ towards $\bar e$.}} Let us fix some (small enough) constant $\delta\in(0,1)$. Since $\triangle$ is open and convex, it is clear that there exists a neighborhood $A$ of $0\times Id$ in $G$ such that $\overline{\cup_{g\in A}g(1-\delta)\triangle}\subseteq \triangle$. We can choose $A$ of the form $A=B(0,r)\times W$ where $W$ is a neighborhood of $Id$ in $SO(3)$ and $r>0$. Also one has that for $\ell_0$ large enough, $\ell_0\triangle$ and $\ell_0g(1-\delta)\triangle$ both belong to $\cR$ by \textbf{(P1)} and \textbf{(P3)}. Thus for any $(g,g')\in A\times G$ and any $\ell\geq\ell_0$, $g'\ell\triangle$ and $g'\ell g(1-\delta)\triangle$ both belong to $\cR$. Using \textbf{(P2)} and \textbf{(A4)}, we infer that 
\begin{equation}
E(g'\ell\triangle)\leq E(g'g\ell (1-\delta)\triangle)+\kappa|(g'\ell\triangle)\setminus (g'g\ell(1-\delta)\triangle)|+|\ell\triangle|\alpha(|\ell\triangle|),
\label{equation_g_sup}
\end{equation}
for any $g'\in G$, $g\in A_\ell:=B(0,r\ell)\times W$ and $\ell\geq\ell_0$. Denoting $g'=(u',R')$ and integrating $g$ over $A_\ell$ in \eqref{equation_g_sup}, we obtain 
\begin{equation}
e_\ell(g') \leq \frac{1}{|A_\ell|}\int_{B(u',r\ell)\times R'W}d\lambda(g)\frac{E(g\ell (1-\delta)\triangle)}{|\ell (1-\delta)\triangle|}+O(\delta)+\alpha(|\ell\triangle|).
\label{estim_sup_av}
\end{equation}
By \textbf{(A3)} and Lemma \ref{translation_invariance_f}
\begin{multline}
\left|\frac{1}{|B(u',r\ell)|}\int_{B(u',r\ell)\times R'W}d\lambda(g)\frac{E(g\ell (1-\delta)\triangle)}{|\ell (1-\delta)\triangle|}-\int_{R'W}\hat{e}(R\ell (1-\delta)\triangle)dR\right|\\
\leq |R'W|\alpha(r\ell),
\end{multline}
and
\begin{multline}
\left||R'W|^{-1}\int_{R'W}dR\, \hat{e}(R\ell (1-\delta)\triangle)-\bar e\right|\leq |W|^{-1}\norm{\hat{e}_{\ell(1-\delta)}-\bar e}_{L^1(SO(3),dR)}
\end{multline}
which converges to 0 when $\ell\to\ii$ as proved in the previous step (here $|R'W|$ denotes the Haar measure of the set $R'W$ in $SO(3)$).
Thus
$$\sup_G \hat{e}_\ell \leq \bar e +\alpha(r\ell)+|W|^{-1}\norm{\hat{e}_{\ell(1-\delta)}-\bar e}_{L^1(SO(3),dR)}+O(\delta)+\alpha(|\ell\triangle|).$$
Since we have already proved that $\lim_{\ell\to\ii}\inf_G \hat{e}_\ell=\bar e$ in Step 3, Theorem \ref{limit_specific} is proved by first passing to the limit as $\ell\to\ii$ and then $\delta\to0$.

\begin{remark}\rm
Adapting the previous arguments, one proves that $\hat{e}_\ell(R)=\hat{e}(R\ell\triangle)$ converges to $\bar e$ uniformly on $SO(3)$.
\end{remark}

\medskip

\noindent{\textbf{Step 5.} \textit{The limit $\bar e$ does not depend on the reference set $\triangle$.}}
Assume that there exists another convex set $\triangle'$ with the same properties \textnormal{\textbf{(P1)}--\textbf{(P3)}}, that $E$ also satisfies \textnormal{\textbf{(A1)}--\textbf{(A5)}} with $\triangle'$, and that $\cM_5=\cM_\triangle\cup\cM_{\triangle'}$. Denote by $\bar e'$ the limit of $E(L\triangle')|L\triangle'|^{-1}$. Applying \eqref{estim_below_average}, we obtain
$$\frac{E(L\triangle')}{|L\triangle'|}\geq e^{\rm av}_\ell-\alpha_1(\ell)-\kappa_1\tilde\eta\left(\frac{\ell}{L|\triangle'|^{1/3}}\right)$$
hence taking two sequences $L_n\to\ii$ and $\ell_n\to\ii$ such that $\ell_n/L_n\to0$, we see that $\bar e'\geq \bar e$. The other inequality is obtained by interchanging $\triangle$ and $\triangle'$.

\subsection{Proof of Theorem \ref{limit_general}: limit for general domains}
Let us fix a sequence $\{\Omega_n\}\subseteq\cR$, such that $|\Omega_n|\to\ii$ and ${\rm diam}(\Omega_n)|\Omega_n|^{-1/3}\leq C$.

\medskip

\noindent{\textbf{Step 1.} \textit{Lower bound.}} We start by proving that
\begin{equation}
\liminf_{n\to\ii}\frac{E(\Omega_n)}{|\Omega_n|}\geq \bar e.
\label{estim_below_Omega}
\end{equation}
This is indeed a simple application of Lemma \ref{lemma_estim_below_average} with $\Omega=\Omega_n\in\cR\subseteq\cM_5=\cM$, $\tilde\eta=\eta$ and $\ell:=|\Omega_n|^{1/6}$. Applying \eqref{estim_below_average}, we obtain
$$\frac{E(\Omega_n)}{|\Omega_n|} \geq e_{|\Omega_n|^{1/6}}^{\rm av}-\alpha_1\left(|\Omega_n|^{1/6}\right) -\kappa_1\eta\left(|\Omega_n|^{-1/6}\right)$$
which proves \eqref{estim_below_Omega} since $e_{|\Omega_n|^{1/6}}^{\rm av}\to \bar e$ by the proof of Theorem \ref{limit_specific}. The rest of the proof is then devoted to showing the upper bound 
\begin{equation}
\limsup_{n\to\ii}\frac{E(\Omega_n)}{|\Omega_n|}\leq \bar e.
\label{estim_above_Omega2}
\end{equation}

\medskip

\noindent{\textbf{Step 2.} \textit{The inner approximation of $\Omega_n$.}} By assumption, we have that the smallest ball $B_n$ containing $\Omega_n$ satisfies $|\Omega_n|/|B_n|\geq \delta'=C^{-3}$ independently of $n$. On the other hand, since $\triangle$ is open, it contains a ball $B(0,r)$, $r>0$. Translating and dilating $\triangle$, we therefore obtain that there exists $\triangle'_n:=\ell'_ng_n\triangle$ for some $\{g_n\}\subset G$ and $\ell'_n\to\ii$ such that $\Omega_n\subseteq \triangle'_n$ for all $n\geq1$  and
$$1\geq\frac{|\Omega_n|}{|\triangle'_n|}\geq c>0.$$

We now consider the tiling $\{\mu\triangle\}_{\mu\in\Gamma}$ of $\R^3$. We fix a sequence $\{\ell_n\}_{n\geq1}$ satisfying $\ell_n/\ell'_n\to0$ and which will be specified below. We apply \textbf{(A6)} with $\Omega=\triangle_n'$ which belongs to $\cR$ by \textbf{(P3)}. 

We notice that there exists a $G$-invariant (with the obvious $G$ action on $G/\Gamma$)
measure $d\hat\lambda$ on $G/\Gamma$ such that 
\begin{equation}
\int f(g) d\lambda(g)=\int_{G/\Gamma}\hat{f}(\sigma) d\hat{\lambda}(\sigma),
\label{formule_quotient}
\end{equation}
where to any function $f\in L^1(G)$ we have defined 
$\hat{f}:G/\Gamma\to \C$ by
$$
\hat{f}([g])=\sum_{\mu\in \Gamma} f(g\mu),
$$
$g$ being any representative for the left coset $[g]=\{g\mu,\mu\in\Gamma\}\in G/\Gamma$. Taking $f(g)=\1_\triangle(gx)$ for some fixed $x\in\R^3$ and using that  $\triangle$ defines a $\Gamma$-tiling, we obtain in particular from \eqref{prop_sliding_group} that
\begin{equation}
\int_{G/\Gamma}d\hat{\lambda}(\sigma)=|\triangle|.
\label{int_G/Gamma}
\end{equation}

Hence \textbf{(A6.3)} can be equivalently written
$$\frac{1}{|\triangle|}\int_{G/\Gamma}d\hat\lambda([g])\sum_{\substack{\mu,\nu\in \Gamma\\ \mu\neq\nu}}I^{\triangle_n'}_{\ell_n}(g\mu,g\nu)\geq -|\triangle_n'|\alpha(\ell_n).$$
We deduce from \eqref{int_G/Gamma} that there exists a rotation and translation $g_n\in G$ of the tiling such that
\begin{equation}
\sum_{\mu,\nu\in\Gamma\atop\mu\ne \nu}I^{\triangle'_n}_{\ell_n}(g_n\mu,g_n\nu)\geq -|\triangle'_n|\alpha(\ell_n)\geq -|\Omega_n|c^{-1}\alpha(\ell_n). 
\label{choice_gn}
\end{equation}
Let us introduce the notation $\triangle_n(\mu):=\ell_ng_n\mu(1+\tau(\ell_n))\triangle$. We define the set of $\mu$'s which are such that the associated $\triangle_n(\mu)$'s are strictly inside $\Omega_n$:
$$\cP_n:=\left\{\mu\in\Gamma\ |\ \triangle_n(\mu)\subset\Omega_n\ \text{ and }\   \text{d}(\triangle_n(\mu),\partial\Omega_n)> \delta\right\}$$
where $\delta$ is the constant appearing in \textbf{(A4)}. We finally introduce the inner approximation of $\Omega_n$ by the tiling (see \eqref{def_approx_tiling_general})
$$A_n=A_{\tau(\ell_n),\ell_n,g_n}(\Omega_n)=\bigcup_{\mu\in\cP_n}\triangle_n(\mu)\subset\Omega_n.$$

By Proposition \ref{reg_inner_approx} and \textbf{(P4)}, we know that $A_n\in\cR'\subseteq\cR_{\eta'}$ for $\ell_n$ large enough (notice that we use $\ell_n/\ell_n'\to0$). We can thus apply \textbf{(A4)} and obtain
\begin{equation}
E(\Omega_n)\leq E(A_n)+\kappa|\Omega_n\setminus A_n|+|\Omega_n|\alpha(|\Omega_n|)=E(A_n)+o(|\Omega_n|)
\label{estim_An}
\end{equation}
and it remains to estimate $E(A_n)$.

\begin{figure}[h]
\centering
\includegraphics[width=10cm]{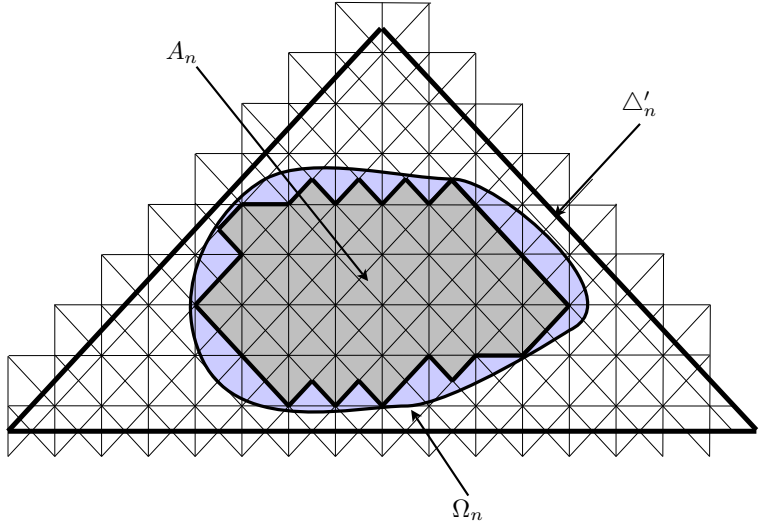}
\caption{The three sets $A_n\subseteq\Omega_n\subseteq\triangle_n'$ and the tiling $\{\triangle_n(\mu)\}_\mu$ used in the proof (with $\tau=0$ here).}
\label{image_preuve}
\end{figure}

\medskip

\noindent{\textbf{Step 3.} \textit{Estimate on $E(A_n)$.}} For the sake of clarity, we introduce the notations $$E_n(\mu)=E^{\triangle_n'}_{\ell_n}(g_n\mu),\quad I_n(\mu,\nu)=I_{\ell_n}^{\triangle'_n}(g_n\mu,g_n\nu)\quad \text{and}\quad s_n(\cP)=s^{\triangle_n'}_{\ell_n}(g_n,\cP).$$
By \textbf{(A6.2)}, we have
\begin{equation}
E(A_n) \leq \sum_{\mu\in\cP_n}E_n(\mu)+\frac12\sum_{\substack{\mu,\nu\in\cP_n\\ \mu\neq\nu}}I_n(\mu,\nu)-s_n(\cP_n)+|A_n|\alpha(\ell_n).
\end{equation}
Then, by \textbf{(A6.1)}, we infer
\begin{align}
E(A_n) &\leq E(\triangle'_n)-\sum_{\mu\in\Gamma\setminus\cP_n}E_n(\mu)-\frac12\sum_{\substack{\mu\in\Gamma,\nu\in\Gamma\setminus\cP_n\\  \mu\neq\nu}}I_n(\mu,\nu)+s_n(\Gamma)-s_n(\cP_n)\nonumber\\
&\qquad\qquad-\frac12\sum_{\substack{\mu\in\Gamma\setminus\cP_n,\nu\in\cP_n}}I_n(\mu,\nu)+2|\triangle'_n|\alpha(\ell_n)\nonumber\\
&\leq E(\triangle'_n)-\sum_{\mu\in\Gamma\setminus\cP_n}E_n(\mu)-\frac12\sum_{\substack{\mu\in\Gamma,\nu\in\Gamma\setminus\cP_n\\  \mu\neq\nu}}I_n(\mu,\nu)\nonumber\\
&\qquad\qquad-\frac12\sum_{\substack{\mu\in\Gamma\setminus\cP_n,\nu\in\cP_n}}I_n(\mu,\nu)+2|\triangle'_n|\alpha(\ell_n)
\label{estim_An_2}
\end{align}
where we have used that $s_n(\Gamma)-s_n(\cP_n)\leq0$ by \eqref{s_monotone}.

Notice by Theorem \ref{limit_specific}, $E(\triangle'_n)=\bar e|\triangle'_n|+o(|\triangle'_n|)$. In order to estimate the terms of the right hand side of \eqref{estim_An_2}, we shall need the following two lemmas, whose proofs will be postponed until the end of the proof of this section.

\begin{lemma}\label{lemmaE}
There exists a function $\alpha_2$ tending to 0 at infinity, such that for any $\cP\subseteq\Gamma$,
\begin{equation}
\left|s_n(\cP)\right|+\left|\sum_{\mu\in\cP}\left(E_n(\mu)-\bar e|\triangle'_n\cap\triangle_n(\mu)|\right)\right|\leq |\triangle'_n|\alpha_2(\ell_n).
\end{equation}
\end{lemma}
\begin{lemma}\label{lemmaI}
There exists a decreasing function $\alpha_3$ tending to 0 at infinity, such that for any $\cP,\cP'\subseteq\Gamma$,
\begin{equation}
\sum_{\substack{\mu\in\cP,\nu\in\cP'\\ \mu\neq\nu}}I_n(\mu,\nu)\geq -|\triangle'_n|\alpha_3(\ell_n)\left(\frac{\ell'_n}{\ell_n}\right)^3.
\end{equation}
\end{lemma}

We now show how to end the proof of Theorem \ref{limit_general} assuming that Lemmas \ref{lemmaE} and \ref{lemmaI} hold. Indeed we deduce from \eqref{estim_An_2} and Theorem \ref{limit_specific} that
\begin{eqnarray}
E(A_n) &\leq& \bar e|\triangle'_n| -\sum_{\mu\in\Gamma\setminus\cP_n}\bar e|\triangle'_n\cap\triangle_n(\mu)|
+|\triangle'_n|\alpha_3(\ell_n)\left(\frac{\ell'_n}{\ell_n}\right)^3\nonumber\\  
& & \qquad \qquad \qquad+|\triangle'_n|(\alpha_2(\ell_n)+2\alpha(\ell_n))+o(|\triangle_n'|)\nonumber\\
 & \leq & \bar e|\Omega_n|+|\triangle'_n|\alpha_3(\ell_n)\left(\frac{\ell'_n}{\ell_n}\right)^3\nonumber\\
 & & \qquad \qquad \qquad+|\triangle'_n|(\alpha_2(\ell_n)+2\alpha(\ell_n))+o(|\triangle_n'|).
\label{estim_An_3}
\end{eqnarray}
Now, we can find a sequence $\{\ell_n\}$ satisfying $\ell_n\to\infty$, $\ell_n/\ell_n'\to0$ and $\alpha_2(\ell_n)(\ell_n'/\ell_n)^3\to0$ as $n\to\ii$. It suffices to take for $\ell_n$ the solution of the equation
$$(\ell'_n)^3=\frac{(\ell_n)^3}{\sqrt{\alpha_3(\ell_n)}}$$
which always exists and satisfies the desired properties since $\alpha_3$ is decreasing and tends to 0 at infinity. Then, the upper bound \eqref{estim_above_Omega2} is a consequence of \eqref{estim_An} and \eqref{estim_An_3} and it only remains to prove Lemmas \ref{lemmaE} and \ref{lemmaI}.

\medskip

\noindent{\textbf{Step 4.} \textit{Proof of Lemmas \ref{lemmaE} and \ref{lemmaI}.}} We shall need a special treatment for the $\triangle_n(\mu)$ which are close to the boundary of $\triangle_n'$. For this reason, we first introduce the following set
$$\cB_n:=\left\{\mu\in\Gamma,\ |\triangle'_n\cap\triangle_n(\mu)|\leq |\triangle_n(\mu)|\zeta(\ell_n)\right\}$$
where $\zeta$ is some function tending to 0 at infinity. Indeed, the following choice will be convenient:
\begin{equation}
\label{def_zeta}
\zeta(\ell_n)=\max\left(\alpha_1\left(|\triangle|^{1/6}\sqrt{\ell_n}\right)^{1/2},|\triangle|^{1/6}\ell_n^{-1/2}\right).
\end{equation}
We start by proving the
\begin{lemma}\label{lemme_estim_below_alpha3} There exists a function $\alpha_4$ tending to 0 at infinity, such that the following holds, for $n$ large enough:
\begin{equation}
\forall \mu\in\Gamma\setminus\cB_n,\ \nu\in\Gamma,\qquad  \frac{E(\triangle'_n\cap(\triangle_n(\mu)\cup\triangle_n(\nu)))}{|\triangle'_n\cap(\triangle_n(\mu)\cup\triangle_n(\nu))|}\geq \bar e -\alpha_4(\ell_n).
\label{estim_below_alpha3a}
\end{equation}
\end{lemma}

\begin{proof}[Proof of Lemma \ref{lemme_estim_below_alpha3}]
We first notice that when $\mu\in\Gamma\setminus\cB_n$ and $\nu\in\Gamma$, then 
\begin{equation}
|\triangle'_n\cap(\triangle_n(\mu)\cup\triangle_n(\nu))| > |\triangle_n(\mu)|\zeta(\ell_n)
\label{condition_Bn}
\end{equation}
by the definition of $\cB_n$. We now prove that $\triangle'_n\cap(\triangle_n(\mu)\cup\triangle_n(\nu))$ satisfies the $\eta_n$-boundary property for the function $\eta_n(t)=c\zeta(\ell_n)^{-2/3}t$ where $c$ is a constant. For the sake of simplicity, we only treat the case where $\mu=\nu\in\Gamma\setminus\cB_n$, the argument being the same for the general case.

Since $\triangle$ is a polyhedron, then $\Theta:=\triangle'_n\cap\triangle_n(\mu)$ is also a polyhedron. Therefore, for any $a>0$, there exists a uniform constant $c'$ such that
\begin{equation}
t/\ell_n\leq a\Longrightarrow |\{x\ |\ \text{d}(x,\partial\Theta)\leq t\}|\leq c' (\ell_n)^2t.
\label{estim_Theta}
\end{equation}
To prove this, it suffices to notice that the boundary of $\Theta$ is composed of a number of faces which is at most twice the number of faces of $\triangle$, and that the area of each of these faces is at most of order $(\ell_n)^2$. If we assume that $t|\Theta|^{-1/3}\leq 1$, then since $|\Theta|\leq |\triangle_n(\mu)|$, we have $t/\ell_n\leq a$ for some $a$ and we can apply \eqref{estim_Theta}. We get that when $t|\Theta|^{-1/3}\leq 1$,
$$|\{x\ |\ \text{d}(x,\partial\Theta)\leq t\}|\leq \frac{c' (\ell_n)^2}{|\triangle_n(\mu)|^{2/3}\zeta(\ell_n)^{2/3}}|\Theta|^{2/3}t\leq c\zeta(\ell_n)^{-2/3}|\Theta|^{2/3}t,$$
where we have used that $|\Theta|>|\triangle_n(\mu)|\zeta(\ell_n)$ since $\mu\notin\cB_n$. This proves that
$\Theta$ satisfies the $\eta_n$-regular property with $\eta_n(t)=c\zeta(\ell_n)^{-2/3}t$, $t\in[0,1)$.

We are now able to apply Lemma \ref{lemma_estim_below_average} with $\Omega=\Theta\subseteq\triangle_n(\mu)$ and $\ell=|\Theta|^{1/6}|\triangle|^{1/6}$. Note that 
$\ell{|\Theta|^{-1/3}}=|\triangle|^{1/6}{|\Theta|^{-1/6}}\leq {\ell_n^{-1/2}\zeta(\ell_n)^{-1/6}}$ which tends to zero as $n\to\ii$ by the definition of $\zeta(\ell_n)$. Hence for $n$ large enough we have $\ell|\Theta|^{-1/3}\leq1/\kappa_1$ and we can apply Lemma \ref{lemma_estim_below_average}.
Denoting $\beta(r)=\bar e-e_r^{\rm av}$ which tends to 0 when $r\to\ii$ by the proof of Theorem \ref{limit_specific}, we obtain
\begin{eqnarray*}
\frac{E(\Theta)}{|\Theta|} & \geq & \bar e -\beta\left(|\Theta|^{1/6}|\triangle|^{1/6}\right)-\frac{|\triangle_n(\mu)|}{|\Theta|}\alpha_1\left(|\Theta|^{1/6}|\triangle|^{1/6}\right)\\
 & &\qquad\qquad-\kappa_1c\frac{\zeta(\ell_n)^{-2/3}|\triangle|^{1/6}}{|\Theta|^{1/6}}\\
 & \geq & \bar e -\beta\left(|\Theta|^{1/6}|\triangle|^{1/6}\right)-\frac{\alpha_1\left(|\triangle|^{1/6}\sqrt{\ell_n}\right)}{\zeta(\ell_n)}-\frac{C|\triangle|^{1/6}}{\zeta(\ell_n)^{5/6}(\ell_n)^{1/2}}.
\end{eqnarray*}
Using the definition \eqref{def_zeta} of $\zeta(\ell_n)$, we obtain
$$\frac{E(\Theta)}{|\Theta|}\geq \bar e -\sup_{t\geq|\triangle|^{1/6}(\ell_n)^{1/2}\zeta(\ell_n)^{1/6}}\{\beta(t)\}-\alpha_1\left(|\triangle|^{\frac16}\sqrt{\ell_n}\right)^{\frac12}-C\big(\ell_n/|\triangle|^{\frac13}\big)^{-\frac1{12}}$$
which ends the proof of Lemma \ref{lemme_estim_below_alpha3}.
\end{proof}

\begin{proof}[Proof of Lemma \ref{lemmaE}] By Theorem \ref{limit_specific}, we have $E(\triangle_n')  =  \bar e|\triangle_n'|+|\triangle_n'|\tilde\alpha(\ell_n')$ for some function $\tilde\alpha$ tending to zero at infinity. By \textbf{(A6.1)} and the choice of $g_n$ to ensure \eqref{choice_gn}, we have
\begin{multline}
\sum_{\mu\in\Gamma}E_n(\mu)-s_n(\Gamma)+|\triangle'_n|\alpha(\ell_n)\leq E(\triangle_n')  \leq  \bar e|\triangle_n'|+|\triangle_n'|\sup_{t\geq\ell_n}\tilde\alpha(t)\\
  \leq  \bar e\sum_{\mu\in\Gamma}|\triangle_n'\cap\triangle_n(\mu)|+|\triangle_n'|\alpha_5(\ell_n),\label{estim_total_Gamma}
\end{multline}
for some function $\alpha_5$ tending to zero at infinity. In the first line of \eqref{estim_total_Gamma} we have used that $\ell_n/\ell_n'\leq1$ (indeed we shall choose $\ell_n/\ell_n'\to0$). In the second line, we have used that $\triangle$ is a polyhedron, hence belongs to $\cR_{\eta_\triangle}$ for some $\eta_\triangle(t)=a|t|$, which easily shows that
\begin{equation}
 \left|\sum_{\mu\in\Gamma}|\triangle_n'\cap\triangle_n(\mu)|- |\triangle_n'|\right|\leq C|\triangle'_n|\tau(\ell_n).
\label{estim_diff_simplices}
\end{equation}
By the nonpositivity \eqref{nonpositivity} of $s_n$, we deduce from \eqref{estim_total_Gamma} that
\begin{equation}
\sum_{\mu\in\Gamma}\left(E_n(\mu) - \bar e|\triangle_n'\cap\triangle_n(\mu)|\right)\leq|\triangle_n'|(\alpha_5(\ell_n)-\alpha(\ell_n)).
\label{estim_total_Gamma2}
\end{equation}

By \textbf{(A6.2)} and \textbf{(A6.5)}, we have for all $\mu\in\Gamma$
$$E_n(\mu)\geq E(\triangle_n'\cap\triangle_n(\mu))-|\triangle_n'\cap\triangle_n(\mu)|\alpha(\ell_n)$$
Let us then consider some $\cP\subseteq\Gamma$. Summing over $\mu\in\cP$ and using Lemma \ref{lemme_estim_below_alpha3} with $\mu=\nu$ and \textbf{(A2)}, we obtain
\begin{eqnarray*}
\sum_{\mu\in\cP}E_n(\mu) & \geq & \sum_{\mu\in\cP\setminus\cB_n}(\bar e -\alpha_4(\ell_n))|\triangle_n'\cap\triangle_n(\mu)|-\kappa\sum_{\mu\in\cP\cap\cB_n}|\triangle_n'\cap\triangle_n(\mu)|\\
 & & \qquad \qquad -\alpha(\ell_n)\sum_{\mu\in\cP}|\triangle_n'\cap\triangle_n(\mu)|\\
 & \geq & \bar e\sum_{\mu\in\cP}|\triangle_n'\cap\triangle_n(\mu)|- (\kappa+\bar e)\sum_{\mu\in\cP\cap\cB_n}|\triangle'_n\cap\triangle_n(\mu)|\\
 & & \qquad\qquad -2(\alpha(\ell_n)+\alpha_4(\ell_n))|\triangle_n'|,
\end{eqnarray*}
where the constant $2$ comes from \eqref{estim_diff_simplices}.
Then, we notice that, for $n$ large enough and by the definition of $\cB_n$
\begin{eqnarray*}
\sum_{\mu\in\cP\cap\cB_n}|\triangle'_n\cap\triangle_n(\mu)| & \leq & \zeta(\ell_n)\sum_{\substack{\mu\in\Gamma\\ \triangle_n(\mu)\cap\partial\triangle_n'\neq\emptyset}}|\triangle_n(\mu)|\\
 & \leq & C\zeta(\ell_n)\left(\frac{\ell'_n}{\ell_n}\right)^2(\ell_n)^3\leq C'|\triangle'_n|\zeta(\ell_n).
\end{eqnarray*}
All this shows that 
\begin{equation}
\sum_{\mu\in\cP}\left(E_n(\mu)-\bar e|\triangle_n'\cap\triangle_n(\mu)|\right)\geq -|\triangle_n'|\alpha_6(\ell_n)
\label{estim_partiel_P}
\end{equation}
for some function $\alpha_6$ independent of $\cP$ and tending to 0 at infinity.

Applying this result to $\cP=\Gamma$ and recalling \eqref{estim_total_Gamma2}, we deduce that
\begin{equation}
\left|\sum_{\mu\in\Gamma}\left(E_n(\mu)-\bar e|\triangle_n'\cap\triangle_n(\mu)|\right)\right|\leq |\triangle_n'|\alpha_7(\ell_n)
\label{estim_partiel_P2}
\end{equation}
with
$$\alpha_7(\ell_n)=|\alpha_5(\ell_n)-\alpha(\ell_n)|+|\alpha_6(\ell_n)|.$$
We can now deduce the upper bound for any $\cP$ by 
\begin{align*}
&\sum_{\mu\in\cP}\left(E_n(\mu)-\bar e|\triangle_n'\cap\triangle_n(\mu)|\right)\\
&\qquad\quad \leq |\triangle_n'|\alpha_7(\ell_n)-\sum_{\mu\in\Gamma\setminus\cP}\left(E_n(\mu)-\bar e|\triangle_n'\cap\triangle_n(\mu)|\right)\\
&\qquad\quad\leq |\triangle_n'|(\alpha_7(\ell_n)+\alpha_6(\ell_n)).
\end{align*}
Also \eqref{estim_partiel_P} and \eqref{estim_total_Gamma} give
$$-s_n(\Gamma)\leq |\triangle_n'|(\alpha_5(\ell_n)-\alpha(\ell_n)+\alpha_6(\ell_n)).$$
Hence by \eqref{nonpositivity} and \eqref{s_monotone},
$$0\leq -s_n(\cP)\leq -s_n(\Gamma)\leq |\triangle_n'|(\alpha_5(\ell_n)-\alpha(\ell_n)+\alpha_6(\ell_n))$$
for any $\cP\subseteq\Gamma$. This ends the proof of Lemma \ref{lemmaE}.
\end{proof}

Before turning to the proof of Lemma \ref{lemmaI}, we quote the following general property of $s^\Omega_\ell$:
\begin{lemma}\label{lemmaT}
Assume that $s^\Omega_\ell:G\times\{\cP\subseteq\Gamma\}\to\R$ satisfies \textnormal{\textbf{(A6.5)}} and \textnormal{\textbf{(A6.6)}}. Then we have for any $\cP\subseteq\Gamma$ with $\#\cP<\ii$
\begin{equation}
 s_\ell^\Omega(g,\cP)\leq \frac{1}{\#\cP}\sum_{\substack{\mu,\nu\in\cP\\ \mu\neq\nu}} s_\ell^\Omega(g,\{\mu,\nu\}).
\label{eq_lemmaT}
\end{equation}
\end{lemma}
We shall give the proof of Lemma \ref{lemmaT} later on and rather turn to the

\begin{proof}[Proof of Lemma \ref{lemmaI}]
Let $\cP,\cP'\subset\Gamma$. By \textbf{(A6.4)}, we have $I_n(\mu,\nu)=0$ if $\triangle_n(\mu)\cap\triangle_n'=\emptyset$ or $\triangle_n(\nu)\cap\triangle_n'=\emptyset$. Hence we can prove Lemma \ref{lemmaI} assuming that $\triangle_n(\mu)\cap\triangle_n'\neq\emptyset$ for all $\mu\in\cP\cup\cP'$. In particular
 \begin{equation}
 \#(\cP\cup\cP')\leq C \left(\frac{\ell'_n}{\ell_n}\right)^3
\label{estim_number_P_P'}
\end{equation}
for some constant $C$.
Then we write
\begin{multline}
\sum_{\substack{\mu\in\cP,\nu\in\cP'\\ \mu\neq\nu}}I_n(\mu,\nu)=
\sum_{\substack{\mu\in\cP\cap\cB_n,\nu\in\cP'\setminus\cB_n\\ \mu\neq\nu}}I_n(\mu,\nu)
+\sum_{\substack{\mu\in\cP\setminus\cB_n,\nu\in\cP'\\ \mu\neq\nu}}I_n(\mu,\nu)\\
+\sum_{\substack{\mu\in\cP\cap\cB_n,\nu\in\cP'\cap\cB_n\\ \mu\neq\nu}}I_n(\mu,\nu).
\label{decomposition_I}
\end{multline}
By \textbf{(A6.2)}, we have for any $\mu,\nu\in\Gamma$,
\begin{multline}
I_n(\mu,\nu)-s_n(\{\mu,\nu\})\geq E\big(\triangle_n'\cap(\triangle_n(\mu)\cup\triangle_n(\nu))\big)-E_n(\mu)-E_n(\nu)\\
-|\triangle_n'\cap(\triangle_n(\mu)\cup\triangle_n(\nu))|\alpha(\ell_n).
\end{multline}
If $\mu,\nu\in\cB_n$, we have by \textbf{(A2)}
\begin{align}
&I_n(\mu,\nu)-s_n(\{\mu,\nu\})\nonumber\\
&\quad \geq -(\kappa+\alpha(\ell_n))\big|\triangle_n'\cap(\triangle_n(\mu)\cup\triangle_n(\nu))\big|-E_n(\mu)-E_n(\nu)\nonumber\\
& \quad  \geq (\bar e|\triangle_n'\cap\triangle_n(\mu)|-E_n(\mu))+(\bar e|\triangle_n'\cap\triangle_n(\nu)|-E_n(\nu))
-C(\ell_n)^{3}\zeta(\ell_n).
\end{align}
Tthus by Lemma \ref{lemmaE}
\begin{align}
&\sum_{\substack{\mu\in\cP\cap\cB_n,\ \nu\in\cP'\cap\cB_n\\ \mu\neq\nu}}I_n(\mu,\nu)\nonumber\\
 & \qquad \geq \sum_{\substack{\mu\in\cP\cap\cB_n,\ \nu\in\cP'\cap\cB_n\\ \mu\neq\nu}}s_n(\{\mu,\nu\})  -2(\#\cB'_n)\alpha_2(\ell_n)|\triangle_n'|-C(\#\cB'_n)^2(\ell_n)^3\zeta(\ell_n)\nonumber\\
 & \qquad \geq\sum_{\substack{\mu\in\cP\cap\cB_n,\nu\in\cP'\cap\cB_n\\ \mu\neq\nu}}s_n(\{\mu,\nu\})  -C'|\triangle'_n|\left(\alpha_2(\ell_n)\left(\frac{\ell'_n}{\ell_n}\right)^2+\frac{\ell'_n}{\ell_n}\zeta(\ell_n)\right),\label{estim_bad_bad}
\end{align}
where we have introduced 
$$\cB'_n:=\cB_n\cap(\cP\cup\cP')$$
and used \textbf{(A6.4)}.

If now $\mu\in\cP\setminus\cB_n$ and $\nu\in\cP'$, for some $\cP,\cP'\subset\Gamma$, $\mu\neq\nu$, we use \textbf{(A6.2)} and Lemma \ref{lemme_estim_below_alpha3} to obtain
\begin{eqnarray*}
I_n(\mu,\nu)-s_n(\{\mu,\nu\}) & \geq &  \bar e|\triangle_n'\cap(\triangle_n(\mu)\cup\triangle_n(\nu))|-E_n(\mu)-E_n(\nu)\\
 & & \qquad -|\triangle_n'\cap(\triangle_n(\mu)\cup\triangle_n(\nu))|(\alpha(\ell_n)+\alpha_4(\ell_n))\\
 & \geq & \big(\bar e|\triangle_n'\cap\triangle_n(\mu)|-E_n(\mu)\big)+\big(\bar e|\triangle_n'\cap\triangle_n(\mu)|-E_n(\mu)\big)\\
 & & \qquad -|\triangle_n'\cap(\triangle_n(\mu)\cup\triangle_n(\nu))|(\alpha(\ell_n)+\alpha_4(\ell_n))\\
 & & \qquad\quad -\alpha_8(\ell_n)|\triangle_n(\mu)|\delta_n(\mu)\delta_n(\nu)
\end{eqnarray*}
for some function $\alpha_8$ tending to 0 at infinity and where we have introduced
$$\delta_n(\mu)=\left\{\begin{array}{rl}
1 &\text{if}\ \triangle_n(\mu)\cap\triangle'_n\neq\emptyset\\
0 &\text{otherwise.}
\end{array}\right.$$
Summing then over $\mu\in\cP\setminus\cB_n$ and $\nu\in\cP'$ and using \textbf{(A6.4)} and Lemma \ref{lemmaE}, we obtain
\begin{equation}
\sum_{\substack{\mu\in\cP\setminus\cB_n,\nu\in\cP'\\ \mu\neq\nu}}I_n(\mu,\nu)\geq \sum_{\substack{\mu\in\cP\setminus\cB_n,\nu\in\cP'\\ \mu\neq\nu}}s_n(\{\mu,\nu\}) -C\left(\frac{\ell'_n}{\ell_n}\right)^3|\triangle_n'|\alpha_9(\ell_n).
\label{estim_bad_good}
\end{equation}
By the nonpositivity of $s_n$ and Lemma \ref{lemmaT}, we have
\begin{multline}
\!\!\!\sum_{\substack{\mu\in\cP\cap\cB_n,\ \nu\in\cP'\setminus\cB_n\\ \mu\neq\nu}}\!\!s_n(\{\mu,\nu\})
+\!\!\!\sum_{\substack{\mu\in\cP\setminus\cB_n,\ \nu\in\cP'\\ \mu\neq\nu}}\!\!s_n(\{\mu,\nu\})
+\!\!\!\sum_{\substack{\mu\in\cP\cap\cB_n,\ \nu\in\cP'\cap\cB_n\\ \mu\neq\nu}}\!\!s_n(\{\mu,\nu\})\\
\geq 3\sum_{\substack{\mu,\nu\in\cP\cup\cP'\\ \mu\neq\nu}}\!\!s_n(\{\mu,\nu\})\geq 3\#(\cP\cup\cP')s_n(\cP\cup\cP')\geq -C\left(\frac{\ell'_n}{\ell_n}\right)^3|\triangle_n'|\alpha_2(\ell_n).
\label{estim_s_below}
\end{multline}
where we have used \eqref{estim_number_P_P'} and Lemma \ref{lemmaE} in the last estimate.
We conclude the proof of Lemma \ref{lemmaI} by \eqref{decomposition_I}, \eqref{estim_bad_bad}, \eqref{estim_bad_good} and \eqref{estim_s_below}.
\end{proof}

We now give the 
\begin{proof}[Proof of Lemma \ref{lemmaT}] For simplicity we use the shorthand notation $s(\cP)=s_\ell^\Omega(g,\cP)$.
We prove \eqref{eq_lemmaT} by induction on $\#\cP$. If $\#\cP=2$, then \eqref{eq_lemmaT} is an identity. Let us assume that \eqref{eq_lemmaT} holds true for any $\cP'\subset\Gamma$ with $\#\cP'\leq N$ and prove it for some $\cP$ with $\#\cP=N+1$.

By the strong subadditivity \textbf{(A6.6)}, we have for all pairs $\{\mu,\nu\}\subset\cP$ with $\mu\neq\nu$
$$s(\cP)\leq s(\cP\setminus\{\mu\})+s(\{\mu,\nu\})\leq \frac{1}{N}\sum_{\substack{\alpha,\beta\in\cP\setminus\{\mu\}\\ \alpha\neq\beta}}s(\{\alpha,\beta\})+s(\{\mu,\nu\}).$$
Now we sum over all pairs $\{\mu,\nu\}\subseteq\cP$:
\begin{equation}
 \frac{N(N+1)}{2}s(\cP) \leq \frac{1}{2N}\sum_{\substack{\mu,\nu\in\cP\\ \mu\neq\nu}}\sum_{\substack{\alpha,\beta\in\cP\setminus\{\mu\}\\ \alpha\neq\beta}}s(\{\alpha,\beta\})+\frac{1}{2}\sum_{\substack{\mu,\nu\in\cP\\ \mu\neq\nu}}s(\{\mu,\nu\}).
\label{sum_over_pairs} 
\end{equation}
Notice 
\begin{eqnarray*}
 \sum_{\substack{\mu,\nu\in\cP\\ \mu\neq\nu}}\sum_{\substack{\alpha,\beta\in\cP\setminus\{\mu\}\\ \alpha\neq\beta}}s(\{\alpha,\beta\})
 & = &  \sum_{\substack{\alpha,\beta\in\cP\\ \alpha\neq\beta}}\sum_{\mu\in\cP\setminus\{\alpha,\beta\}}\sum_{\nu\in\cP\setminus\{\mu\}}s(\{\alpha,\beta\})\\
 & = & N(N-1)\sum_{\substack{\alpha,\beta\in\cP\\ \alpha\neq\beta}}s(\{\alpha,\beta\}).
\end{eqnarray*}
Inserting in \eqref{sum_over_pairs} we obtain
$$ \frac{N(N+1)}{2}s(\cP) \leq \left(\frac{N(N-1)}{2N}+\frac{1}{2}\right)\sum_{\substack{\alpha,\beta\in\cP\\ \alpha\neq\beta}}s(\{\alpha,\beta\})=\frac{N}{2}\sum_{\substack{\alpha,\beta\in\cP\\ \alpha\neq\beta}}s(\{\alpha,\beta\})$$
which proves \eqref{eq_lemmaT} for $\#\cP=N+1$.
\end{proof}

\subsection{Proof of Proposition \ref{reg_inner_approx}: regularity of inner approximation}\label{proof_reg_inner_approx}
Let $\Omega$ be a  fixed set in $\cR_\eta$, $g\in G$, $\tau\in[0,\tau_0]$ and define
\begin{equation}
A=\bigcup_{\substack{\mu\in\Gamma\ |\ \triangle(\mu)\subset\Omega,\\ \text{d}(\partial\triangle(\mu),\partial\Omega)>\delta}}\triangle(\mu)\quad \text{with}\quad \triangle(\mu)=\ell g\mu(1+\tau)\triangle.
\label{def_approx_tiling_general2}
\end{equation}
We also introduce $\cP:=\{\mu\in\Gamma\ |\ \triangle(\mu)\subset A,\ \triangle(\mu)\cap\partial A\neq\emptyset\}$.
We have to estimate $\left|\left\{x\ |\ \text{d}(x,\partial A)\leq |A|^{1/3}t\right\}\right|$. 
Let us assume first that $t\leq |\triangle(\mu)|^{1/3}|A|^{-1/3}c$. In this case, we write
$$\left\{x\ |\ \text{d}(x,\partial A)\leq |A|^{1/3}t\right\}\subseteq \bigcup_{\mu\in\cP}\left\{x\ |\ \text{d}(x,\partial \triangle(\mu))\leq |A|^{1/3}t\right\}$$
and, using that $\triangle(\mu)\in\cR_\eta$, we infer
\begin{equation}
\left|\left\{x\ |\ \text{d}(x,\partial A)\leq |A|^{1/3}t\right\}\right|\leq (\#\cP)|\triangle(\mu)|\times\eta\left(\frac{t|A|^{1/3}}{|\triangle(\mu)|^{1/3}}\right)
\label{estim_Omega_tiling1}
\end{equation}
(recall that $\eta$ is by assumption defined on $[0,c)$).
If we assume that $\ell$ is large enough compared to $\delta$, $\ell\geq\ell_0$, then there exists a constant $\gamma$ such that
$$\forall\mu\in\cP,\qquad \triangle(\mu)\subseteq \left\{x\ |\ \text{d}(x,\partial\Omega)\leq \gamma|\triangle(\mu)|^{1/3}\right\}.$$
In each $\triangle(\mu)$, we can choose a ball $B(\mu)$ of volume proportional to that of $\triangle(\mu)$ such that all the $B(\mu)$, $\mu\in\cP$ never overlap. This implies
$$(\#\cP)|\triangle(\mu)|\leq \gamma' \left|\left\{x\ |\ \text{d}(x,\partial\Omega)\leq \gamma|\triangle(\mu)|^{1/3}\right\}\right|$$
for some constant $\gamma'$. 
Using now that $\Omega$ has an $\eta$-regular boundary, we have
\begin{equation}
\left|\left\{x\ |\ \text{d}(x,\partial\Omega)\leq \gamma|\triangle(\mu)|^{1/3}\right\}\right|\leq |\Omega|\times\eta\left(\frac{\gamma|\triangle(\mu)|^{1/3}}{|\Omega|^{1/3}}\right)
\label{estim_Omega_tiling2}
\end{equation}
when $\gamma|\triangle(\mu)|^{1/3}|\Omega|^{-1/3}\leq c$ i.e. when $|\triangle(\mu)|^{1/3}|\Omega|^{-1/3}$ is small enough, $\ell\leq \delta_0|\Omega|^{1/3}$. Hence 
$$(\#\cP)|\triangle(\mu)|\leq\gamma'{|\Omega|}\times\eta\left(\frac{\gamma|\triangle(\mu)|^{1/3}}{|\Omega|^{1/3}}\right).$$
Inserting in \eqref{estim_Omega_tiling1}, we deduce that
\begin{equation}
\left|\left\{x\ |\ \text{d}(x,\partial A)\leq |A|^{1/3}t\right\}\right|\leq \gamma'
|\Omega|\eta\left(\frac{t|A|^{1/3}}{|\triangle(\mu)|^{1/3}}\right)\eta\left(\frac{\gamma|\triangle(\mu)|^{1/3}}{|\Omega|^{1/3}}\right)
\label{estim_Omega_tiling3}
\end{equation}
when $t\leq|\triangle(\mu)|^{1/3}|A|^{-1/3}c$.
By the same arguments as above, it can be proved that 
$|\Omega|\leq |A|+C|\Omega|\eta\left(\gamma|\triangle(\mu)|^{1/3}|\Omega|^{-1/3}\right),$
i.e. we have for instance $|\Omega|\leq2|A|$. We now use the specific form of $\eta\in\cE$, $\eta(u)=a|u|^b$ to infer
\begin{equation}
\left|\left\{x\ |\ \text{d}(x,\partial A)\leq t|A|^{1/3}\right\}\right|\leq 2\gamma'a\gamma^b|A|\eta\left(t\right).
\label{premiere_estimation_frontiere2}
\end{equation}

If on the other hand $t\geq c|\triangle(\mu)|^{1/3}|A|^{-1/3}$, then we just say directly that
\begin{eqnarray*}
\left\{x\ |\ \text{d}(x,\partial A)\leq t|A|^{1/3}\right\} & \subseteq & \left\{x\ |\ \text{d}(x,\partial\Omega)\leq \gamma|\triangle(\mu)|^{1/3}+t|A|^{1/3}\right\}\\
 & \subseteq &\left\{x\ |\ \text{d}(x,\partial\Omega)\leq (c+\gamma)t|A|^{1/3}\right\}
\end{eqnarray*}
and thus for $t\leq c(c+\gamma)^{-1}$
\begin{equation}
\left|\left\{x\ |\ \text{d}(x,\partial A)\leq t|A|^{1/3}\right\}\right|\leq 2|A|(1+\gamma)^b\eta\left(t\right).
\label{deuxieme_estimation_frontiere}
\end{equation}
As a conclusion, $A\in\cR_{\tilde\eta}$ for $\tilde\eta(t)=m\eta(t)$ on $[0,c/m)$ and with $m=\max\left(2(1+\gamma)^b,2\gamma'a\gamma^b,c+\gamma\right)$.
\qed

\subsection{Proof of Proposition \ref{A6_implies_A5}: \textbf{(A6)} implies \textbf{(A5)}}\label{sec:proof_A5_implies_A6}
For any $\ell'>0$ we denote by $\ell=f(\ell')$ the smallest positive number such that $\ell'=\ell(1+\tau(\ell))$. Notice that $\lim_{\ell'\to\ii}f(\ell')=\ii$ since $\ell(1+\norm{\tau}_{L^\ii})\geq \ell'$. For any fixed $\ell'$, we apply (A6) with the corresponding $\ell=f(\ell')$.

Note that the right hand side of \eqref{lower_bound_Omega} indeed only depends of $[g]\in G/\Gamma$ and not $g\in G$. 
Hence we can integrate \textbf{(A6.1)} over $G/\Gamma$ and, using both \eqref{int_G/Gamma}, \textbf{(A6.2)} and \textbf{(A6.3)}, we obtain
\begin{multline*}
E(\Omega)  \geq  \frac{1}{|\triangle|}\int_{G/\Gamma}d\hat{\lambda}([g])\sum_{\mu\in\Gamma}E\big(\Omega\cap\ell g\mu (1+\tau(\ell))\triangle\big) - 3|\Omega|_{\rm r}\alpha(\ell)\\
  -\alpha(\ell)\frac{1}{|\triangle|}\int_{G/\Gamma}d\hat{\lambda}([g])\sum_{\mu\in\Gamma}\big|\Omega\cap\ell g\mu(1+\tau(\ell))\triangle\big|.
\end{multline*}
We then use \eqref{formule_quotient} to infer
\begin{multline*}
\frac{1}{|\triangle|}\int_{G/\Gamma}d\hat{\lambda}([g])\sum_{\mu\in\Gamma}\big|\Omega\cap\ell g\mu (1+\tau(\ell))\triangle\big|  =  \frac{1}{|\triangle|}\int_G\big|\Omega\cap\ell g(1+\tau(\ell))\triangle\big|d{\lambda}(g)\\
 =  \frac{1}{|\ell\triangle|}\int_G\big|\Omega\cap g\ell(1+\tau(\ell))\triangle\big|d{\lambda}(g) =  |\Omega|(1+\tau(\ell))^3
\end{multline*}
by \eqref{eq_volume} and similarly
\begin{multline*}
\frac{1}{|\triangle|}\int_{G/\Gamma}d\hat{\lambda}([g])\sum_{\mu\in\Gamma}E\big(\Omega\cap\ell g\mu (1+\tau(\ell))\triangle\big)   =  \frac{1}{|\ell\triangle|}\int_G E\big(\Omega\cap g\ell'\triangle\big)d{\lambda}(g)\\
 \geq  \frac{1}{|\ell'\triangle|}\int_G E\big(\Omega\cap g\ell'\triangle\big)d{\lambda}(g)
   -\kappa\big((1+\tau(\ell))^3-1\big)|\Omega|,
\end{multline*}
where we have used \textbf{(A2)}. This easily proves \textbf{(A5)}.\qed

\subsection{Proof of Lemma \ref{regularity_convex}: regularity of open convex sets}\label{sec_proof_convex}
Let us fix some open and bounded convex set $K$, containing 0. We have
$$|\{x\in\R^3\ |\ {\rm d}(x,\partial K)\leq t\}|=|\overline{K}+tB|-|\overline{K}\sim tB|$$
where $B$ is the closed unit ball and where we recall that $A\sim B:=\bigcap_{b\in B}(A-b)$, see \cite[Chap. 3]{Schneider}. As $K$ is open and contains 0, there exists a $r>0$ such that $rB\subseteq K$. Hence we have for all $0\leq t\leq r$
$$\overline{K}+tB\subseteq (1+t/r)\overline{K}\quad\text{and}\quad \overline{K}\sim tB\supseteq \overline{K}\sim (t/r)\overline{K}=(1-t/r)\overline{K}.$$
Therefore
\begin{eqnarray*}
 |\{x\in\R^3\ |\ {\rm d}(x,\partial\Omega)\leq t\}| & \leq & |(1+t/r)\overline{K}|-|(1-t/r)\overline{K}|\\
 &= &\left( (1+t/r)^3-(1-t/r)^3\right)|{K}|\leq \frac{8t}{r}|K|.
\end{eqnarray*}
\qed

\addcontentsline{toc}{section}{References}
\bibliographystyle{amsplain}

\end{document}